\documentclass[a4paper, superscriptaddress, amsfonts, amssymb, amsmath, reprint, showkeys, nofootinbib, twoside]{revtex4-1}
\usepackage[english]{babel}
\usepackage[utf8]{inputenc}
\usepackage[colorinlistoftodos, color=green!40, prependcaption]{todonotes}
\usepackage{amsthm}
\usepackage{mathtools}
\usepackage{physics}
\usepackage{xcolor}
\usepackage{graphicx}
\usepackage[left=23mm,right=13mm,top=35mm,columnsep=15pt]{geometry} 
\usepackage{adjustbox}
\usepackage{placeins}
\usepackage[T1]{fontenc}
\usepackage{lipsum}
\usepackage{csquotes}
\usepackage{float}
\usepackage{multirow}
\usepackage{chngpage}
\usepackage{amsmath}
\usepackage{color}

\usepackage{xr}
\makeatletter

\newcommand*{\addFileDependency}[1]{
\typeout{(#1)}
\@addtofilelist{#1}
\IfFileExists{#1}{}{\typeout{No file #1.}}
}\makeatother

\newcommand*{\myexternaldocument}[1]{%
\externaldocument{#1}%
\addFileDependency{#1.tex}%
\addFileDependency{#1.aux}%
}

\myexternaldocument{supplementarymat}

\newcommand{\angstrom}{\textup{\AA}}
\usepackage[pdftex, pdftitle={Article}, pdfauthor={Author}]{hyperref} 
\begin{document}
\title{Alloy Informatics through Ab Initio Charge Density Profiles: Case Study of Hydrogen Effects in Face-Centered Cubic Crystals}

\author{Dario Massa}
    \affiliation{NOMATEN Centre of Excellence, National Centre for Nuclear Research, Warsaw, Poland}
\author{Efthimios Kaxiras}
    \affiliation{Department of Physics, Harvard University, Cambridge, MA, USA}
\author{Stefanos Papanikolaou}
    \email[]{stefanos.papanikolaou@ncbj.gov.pl}
    \affiliation{NOMATEN Centre of Excellence, National Centre for Nuclear Research, Warsaw, Poland}
\date{\today} 

\begin{abstract}
Materials design has traditionally evolved through trial-error approaches, mainly due to the non-local relationship between microstructures and properties such as strength and toughness. We propose 'alloy informatics' as a machine learning based prototype predictive approach for alloys and compounds, using electron charge density profiles derived from first-principle calculations. We demonstrate this framework in the case of hydrogen interstitials in face-centered cubic crystals, showing that their differential electron charge density profiles capture crystal properties and defect-crystal interaction properties. Radial Distribution Functions (RDFs) of defect-induced differential charge density perturbations highlight the resulting screening effect, and, together with hydrogen Bader charges, strongly correlate to a large set of atomic properties of the metal species forming the bulk crystal. We observe the spontaneous emergence of classes of charge responses while coarse-graining over crystal compositions. Nudge-Elastic-Band calculations show that RDFs and charge features also connect to hydrogen migration energy barriers between interstitial sites. Unsupervised machine-learning on RDFs supports classification, unveiling compositional and configurational non-localities in the similarities of the perturbed densities. Electron charge density perturbations may be considered as bias-free descriptors for a large variety of defects. 
\end{abstract}


\maketitle

\section{Introduction} \label{sec:outline}

Hydrogen has, nowadays, a dual importance, both for renewable energy solutions  ~\cite{abe2019hydrogen,moradi2019hydrogen}, as well as for the fundamentals of hydrogen-embrittlement effects  ~\cite{li2020review,GONG2022117488,PhysRevLett.94.155501,djukic2019synergistic,book_embrit_1}. Strict criteria for hydrogen storage systems automotive applications~\cite{DoE} on capacities, working temperatures, reversibility of cycles, safety and price require further improvements in the deployed materials, beyond hydrides  ~\cite{hydrides1,hydrides2,hydrides3} and other solutions. For constrained optimization problems like these, the use of Machine Learning (ML) techniques appears essential for addressing the required complexity and navigating the optimal compositional space. 

Indeed, in materials design and development, the role of ML~\cite{ML1,ML2, 2018learning} methods has been established as crucial in the context of data-driven approaches. These types of methods have evolved into a new field of computational materials design, referred to as Materials Informatics ~\citep{MI_1,MI_2,MI_3,MI_4,MI_5}. Existing open-access simulation and experiment datasets, have been used to develop ML models and predictions for electrocatalysts materials~\cite{electrocatML_1,electrocatML_2}, carbon-capture materials~\cite{CcaptureML_1,CcaptureML_2}, Metal Oxide Frameworks (MOFs)~\cite{MOFs}, organic materials~\cite{organic,organic_2}, and many other classes,  with the common aim of maximizing the screening efficiency and materials performances. Conventionally, the foremost challenge, besides the presence of an efficient and well-tested ML algorithm, is the specification of the set of effective system descriptors.

In the context of ML approaches with atomistic input, i.e. coming from simulation data, such as density functional theory (DFT)  \cite{dft1_general,dft2_general,dft3_specific_catalysis,dft4_general,dft5_specific_alloys, dft6_specific_detection,dft7_batteries}, a large variety of descriptors have been proposed, including fixed-length feature vectors of  elemental or electronic properties~\cite{fixlen_feat_1,fixlen_feat_2,fixlen_feat_3}, as well as structural descriptors, based on rotationally and translationally invariant functions of atomic coordinates, like the Coulomb matrix~\cite{CoulombMat}, the atom-centered symmetry functions (ACSFs)~\cite{ACSFs}, the social permutation invariant coordintes (SPRINT)~\cite{SPRINT}, the smooth overlap of atomic positions (SOAP)~\cite{SOAP} and the global minimum of root mean-square distance~\cite{globalmin}. However, given that the number of DFT calculations is naturally limited, the challenge in selecting the appropriate set of descriptors stems from the necessity for them to effectively encompass element-specific microstructural features while also accounting for potential electron-electron correlations across various length scales.
\begin{figure*}[htb]
    \centering
    \includegraphics[width=0.85\textwidth]{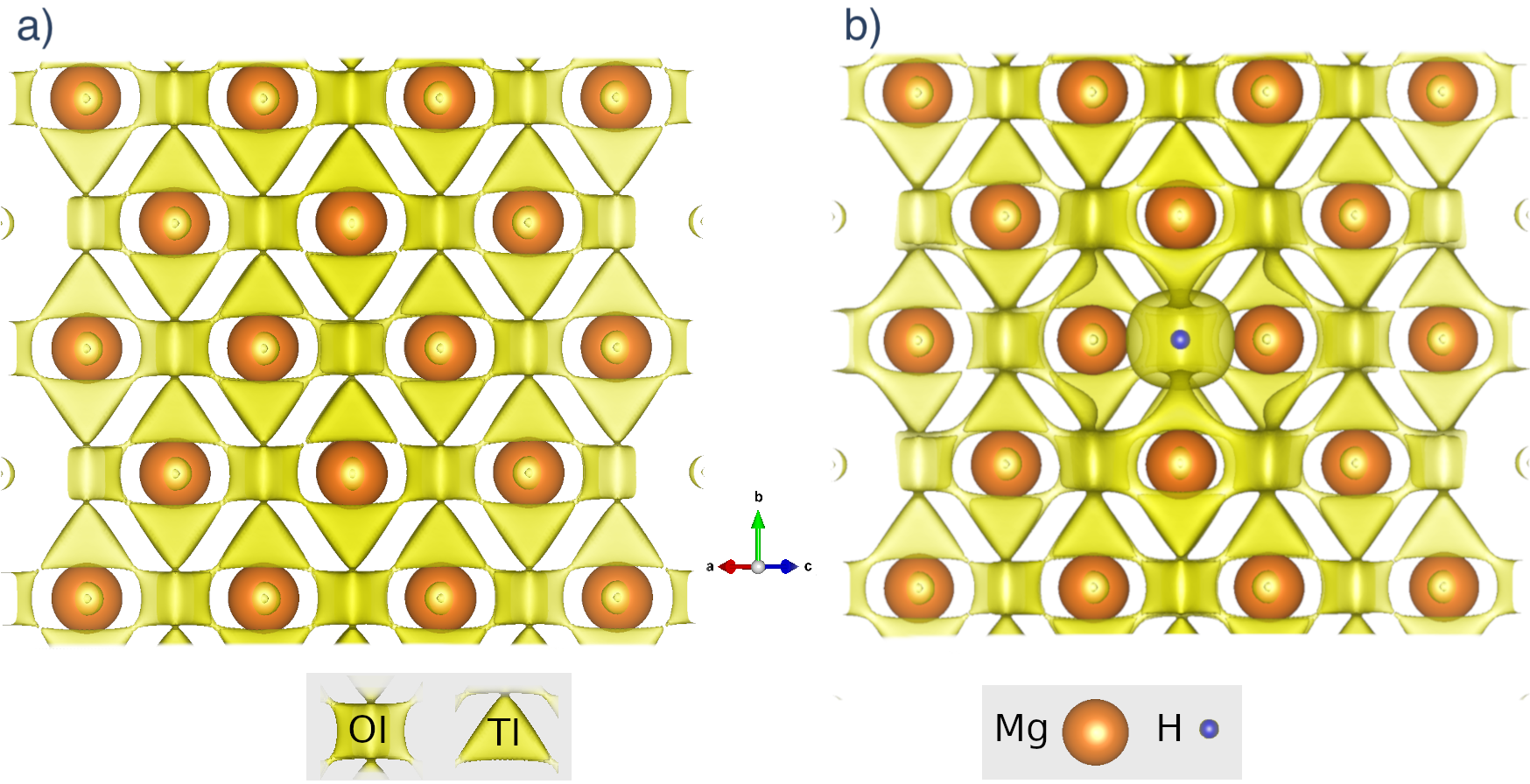}
    \caption{Three dimensional differential electron charge accumulation isolines in a bulk Mg supercell, (a) in the absence of H and  (b) in the presence of an H atom sitting at a high symmetry site. The isoline level, of $0.00183\text{ e/bohr}^3$ in both visualizations, has been set to visibly highlight the two high-symmetry interstitial regions in which hydrogen could stably sit, namely OI and TI.}
    \label{fig:Mgexample}
\end{figure*}

In the search of hydrogen-based energy solutions, as well as in studies of embrittlement effects, hydrogen may be regarded as a defect inside the host material, thus the classification of fundamental interactions among the two for constructing optimal ML descriptors represents the critical challenge. 
In the present work, we use the defect-induced electron charge density perturbations in order to illuminate classification features of local crystalline defects and their properties. We consider hydrogen interstitial defects in a variety of bulk face-centered cubic (FCC) metal crystals as a key demonstrative example of a high-throughput procedure~\cite{HighTrough_1,HighTrough_2}, where our descriptors are constructed on the base of hydrogen differential electron charge density profiles, and of their local quantities, such as radii and extrema. In particular, in Sec.(\ref{subsec:RDF}) we show how the building of a Radial Distribution Function (RDF) in the density fields allows for a fair comparison between the profiles, and the distinctive features of the effects of hydrogen are discussed in terms of charge density perturbations. In Sec.(\ref{subsec:correlations}), we prove that the electron density related features show remarkable correlations with both atomic and structural features of the host bulk crystals, and that they can also be exploited to infer hydrogen mobility in the surrounding matrix. After the building of a density dataset, in Sec.(\ref{subsec:densitydataset}) we perform unsupervised ML, while in Sec.(\ref{subsec:universal}) we investigate over the possibility of a mean-field like treatment of the density perturbations, averaging out compositional peculiarities. Basing on preliminary tests, we discuss interesting trends between the profile features and the properties of the embedding crystals, uncovering compositional relationships (among different crystals) and non-compositional ones (among different hydrogen interstitial equilibrium positions in the same crystals).
The results of this study might pave the way to novel material informatics methods for prediction of defects behaviour in materials and their related effects, by interpretation of the perturbed differential charge density profiles (PChDP) in its surroundings.

\section{Methods} \label{sec:Methods}

We focus on the hydrogen-induced changes in the differential~\cite{footnote_differential} electron charge density profile of an otherwise periodic crystal structure. For this aim, we develop $35$ simple FCC supercells of pure metallic crystals, listed in Tab.S.(I) of the associated Supplementary Material (SM), in which a hydrogen atom is inserted. An example of such considered structures is contained in Fig.(\ref{fig:Mgexample}).

\subsection{Density Functional Theory Methods}
\label{subsec:meth1}

We perform Density-Functional-Theory (DFT) simulations in the \texttt{QUANTUM ESPRESSO} (QE) package \cite{QE1,QE2,QE3} of $2\times2\times2$ supercells of pure FCC metallic bulk crystals, whose list is collected in Tab.S.(I) of SM. Pseudo-Potentials for all  involved atomic species are taken to be Ultra-Soft with an additional implementation of the Perdew-Burke-Ernzerhof (PBE) \cite{PBE} functional. 

In order to better model dispersion forces in the possible presence of weakly interacting states, the non-local exchange-correlation density functional rvv10~\cite{rvv10} from Vydro and Van Voorhis has been also implemented. Further, smearing~\cite{Smearing} has been introduced to correctly deal with metals, and the calculations have been set to be spin-polarized for possible non-trivial magnetic effects. The convergence with respect to the total energy on the number of k-points, plane-waves cutoff energy and smearing spreading is checked, for each case, after a preliminary variable-cell relaxation of the pure metals, and the final equilibrium bulk structures are then obtained via a further fixed-cell relaxation with the optimal parameters. The common acceptance threshold in the variation of the total energy upon parameter change is set to $10^{-5}$ Ry. This process allows to ensure the correct optimization of both parameters and structures for each specific case, which might need different optimal parameters sets. The parameter sets for each atomic species are contained in Tab.S.(I) of SM. The forces and total energy convergence thresholds for ionic minimization are set to a common value of $10^{-5}\text{a.u.}$ and $10^{-6}\text{a.u.}$ respectively. The structural properties like the bulk, Young and shear modulii and the Poisson ratios of the pure optimized samples are then extracted via the \texttt{THERMO\_PW} \cite{thermopw} driver of QE, and are reported in Tab.S.(II) of SM, together with the crystals lattice constants and other relevant structural information resulting from this optimization process. 

In the relaxed, converged crystalline supercells, we insert one hydrogen atom, and let a relaxation minimize the forces in the system, confirming the well known two possible equilibrium interstitial positions of hydrogen in FCC structures: either as an octahedral interstitial (OI) or a tetrahedral interstitial (TI)~\cite{OITI_H}.

In Fig.(\ref{fig:Mgexample}), there is a demonstration of differential electron charge density isolines in the absence and presence of H in a Mg supercell. The accumulation isoline level has been set to the same one in both cases, and in such a way to visibily highlight the two high-symmetry states interstitial regions in which hydrogen could stably sit.

We extract the PChDP around hydrogen atoms, by subtracting the differential density of pure bulk structures from the ones of defected samples, both for OI and TI cases. At the same time, for all samples and configurations, the electron (pseudo-)charge densities and the all-electron ones are also stored and used in Bader charge analysis~\cite{bader}; the latter provides useful information regarding Bader charge exchanges, volumes and radii~\cite{footnote_bader}. In Tab.S.(III) and Tab.S.(IV) of SM, these quantities are documented for each studied crystal. In particular, for building a dataset with atomic, structural and density properties, we consider a set of elemental properties from the \texttt{Mendeleev} package, a Python resource \cite{mendeleev2014}, and we maintain Bader Analysis~\cite{bader} quantities related to the hydrogen atom and to one of its first nearest neighbour. Technically, in an FCC crystal in which hydrogen lives as OI, there are six nearest neighbours, but the extracted Bader information is nearly identical for symmetry. For simplicity and consistency, once the choice of which metal atom should be selected for its Bader properties is done in the OI case, the same metal atom is chosen for the TI case as well.

In order to compute the hydrogen binding energies (BEs) in different crystals, we consider the following formula
\begin{equation}
    E_{\text{BE}} = E_{\text{MH}} - E_{\text{M}} - \dfrac{E_{\text{H2}}}{2}
\end{equation}
where $ E_{\text{MH}}$ is the total energy of the bulk supercell with the hydrogen interstitial atom,   $E_{\text{M}}$ the total energy of the supercell without hydrogen, and $E_{\text{H2}}$ the energy resulting from a fixed-cell relaxation of an hydrogen molecule in a sufficiently large cubic simulation box ($a=24 \angstrom$). The BE is computed with the same QE parameters used for the specific metal bulk system each time in consideration, for computational consistency.

\subsection{Isolines and materials' comparisons}\label{subsec:meth2}
In order to access to a more complete description of the effects of a systematic change in composition of the embedding FCC matrices, and the dependence on the different hydrogen interstitial positions, we investigate the contours of the PChDP around hydrogen. All the density grids are brought to an identical resolution by using~\texttt{mp-pyRho}\cite{mp-pyrho}. Then, identifying a unique radius $R$ in the unit cell to which each value of the density profile can be associated, a Radial Distribution Function (RDF) is calculated \cite{RDF}, representing the basis of our comparison between different profiles. The task is parallelized using the \texttt{Joblib} Python package. Finally, shifting the obtained signals by the hydrogen position in the grid is a key step for efficiently comparing the profiles. In Tab.(\ref{tab:charge_features}) are reported the relevant density-related properties which have been further considered. 

\begin{table}[htb]
    \centering
    \begin{tabular}{c|c}\hline
         Feature & Description \\ \hline
         Ch(M) ex.($\text{e}^{-}$)&exchanged charge from the M atom \\
         Ch(H) ex.($\text{e}^{-}$)& exchanged charge from the H atom \\
         R(M)(\AA)& Bader radius of the M charge \\
         R(H)(\AA)& Bader radius of the H charge \\
         V(M)(\AA$^3$)& Bader volume of the M charge \\
         V(H)(\AA$^3$)& Bader volume of the H charge \\
         RDF peak width& half-height width of the RDF peak \\
         RDF MAX& maximum of the RDF peak \\ \hline
    \end{tabular}
    \caption{Density-related properties promoted as describing features of hydrogen interaction in different materials. (M) stands for Metal and (H) stands for Hydrogen. }
    \label{tab:charge_features}
\end{table}

\subsection{Validation strategies and procedures}\label{subsec:meth3}

\subsubsection{Diffusive properties}\label{subsub:meth3.1}
The key aspects related to hydrogen for industrial applications, especially for renewable energy such as hydrogen storage or catalysis of energy reactions, ultimately rely on its diffusive characteristics in different materials~\cite{h_diff_1,h_diff_2,h_diff_3,h_diff_4}. In order to identify validation routes based on the natural relationship between the extension of a charge profile and its mobility in the surrounding environment, we utilize DFT Nudge-Elastic-Band (NEB) calculations to quantify energy barriers for hydrogen migration along highly-symmetric sites in  bulk metals. In particular, $2\times2\times2$ supercells with one hydrogen atom have been considered as starting and ending images, and the total number of points to discretize the path has been set to $14$, with quasi-Newton Broyden optimization scheme and optimization step length of $1.0 \text{ a.u.}$. The threshold controlling the error (the norm of the force orthogonal to the path) in the climbing-image simulation is set to  $0.05 \text{ eV}/\angstrom$. 

\subsubsection{Size effects in charge density calculations}\label{subsub:meth3.2}
The charge density perturbations due to an interstitial hydrogen defect shall display size effects, that may gain crucial importance for the possibility of undesired artifacts from spurious interactions in small supercells. As sanity checks, in Fig.S.(1) and Tab.S.(V) of the SM we report the comparison of a three dimensional perturbation induced by an hydrogen interstitial defect for a $2\times2\times2$ and $3\times3\times3$ Mo supercell, proving that in typical metals, and thus, for the aims of this study, the expected size effects are negligible. 

\subsubsection{Effects of different crystalline symmetries in charge density considerations}\label{subsub:meth3.3}
For the applicability of the proposed method, we mainly consider the important example of FCC pure bulk crystals, materials that are predominantly metals. Clearly, not all the considered elements form FCC crystals at Standard Temperature and Pressure (STP) conditions, but we opted for a single-phase study for the sake of a simpler comparison among the profiles, avoiding possible differences arising from different crystalline backgrounds and interstitial positions which could complicate the origin of charge density fluctuations. However, the methods discussed in this manuscript are general and shall become applicable to other crystalline structures. For this purpose, the SM contain examples of hydrogen-induced PChDP for non-metallic HCP supercells, including C, Si and Hg, and we promote specific relevant considerations and comparisons with the results presented in the main text. 


\subsection{Machine Learning}\label{subsec:meth4}

\subsubsection{Principal Component Analysis (PCA)}\label{subsec:meth4.1}
In the present work, we follow the Principal Component Analysis \cite{PCA,PCA2} method, probably the most popular choice for developing unsupervised learning, allowing the simultaneous preservation of as much information (variability) as possible. Let us consider a dataset represented by a $\text{n}\times\text{p}$ matrix, $\text{\textbf{D}}$, where $\text{n}$ is the number of samples and $\text{p}$ is the number of features in the dataset. If we define a linear combination of features as \begin{equation}\label{eq:PCA}
    \sum_{j=1}^p\,\,c_j \text{\textbf{d}}_j=\text{\textbf{D}}\text{\textbf{c}}
\end{equation} where $\text{\textbf{c}}$ is a p-dimensional vector of constants, the aim is to search the combination such that its variance is maximized, which means maximizing the quadratic form \begin{equation}\label{eq:covmat}
    var(\text{\textbf{D}}\text{\textbf{c}})=\text{\textbf{c}}^T\text{\textbf{Sc}}
\end{equation}
where \text{\textbf{S}} is the $p \times p$ covariance matrix. Requiring with a Lagrange multiplier the vectors of constants to have unit norm, the actual problem to solve becomes the following\begin{equation}\label{eq:lagrange}
    max\big[ \text{\textbf{c}}^T\text{\textbf{Sc}}-\lambda(\text{\textbf{c}}^T\text{\textbf{c}}-1) \big]
\end{equation} 
and the consequent differentiation with respect to the vector of constants leads to an eigenvalue problem which is at the base of PCA \begin{equation}\label{eq:eigen}
    \text{\textbf{Sc}}=\lambda\text{\textbf{c}}
\end{equation}
Therefore, $\lambda$ are the eigenvalues of the covariance matrix $\text{\textbf{S}}$ corresponding to the unit norm eigenvectors $\text{\textbf{c}}$. Being \text{\textbf{S}} a squared real symmetric matrix, it has exactly $\text{p}$ eigenvalues, corresponding to eigenvectors $\text{\textbf{a}}_k$ that are the solution of the problem of finding $\text{p}$ new linear uncorrelated combinations (Principal Components, PCs) $d_{\text{\textbf{a}}_k}=\sum_{j=1}^p a_{jk}d_j$ that maximise the variance \cite{PCA3}. In a geometric interpretation of PCA, we can interpret the eigenvalues of the covariance matrix as associated to the magnitude of the variance captured by the corresponding PC (axis) in the multivariate cloud space, and the eigenvectors as the orientation of the PC (axis). The so-called loadings, eigenvectors scaled up by variences, are the carriers of the combined information on the captured variance and direction of the PCs, and, representing the covariances between the latter and the original dataset variables, tell us how much of the variation in a variable is explained by the component.

\begin{figure*}[tbh]
    \centering
    \includegraphics[width=0.85\textwidth]{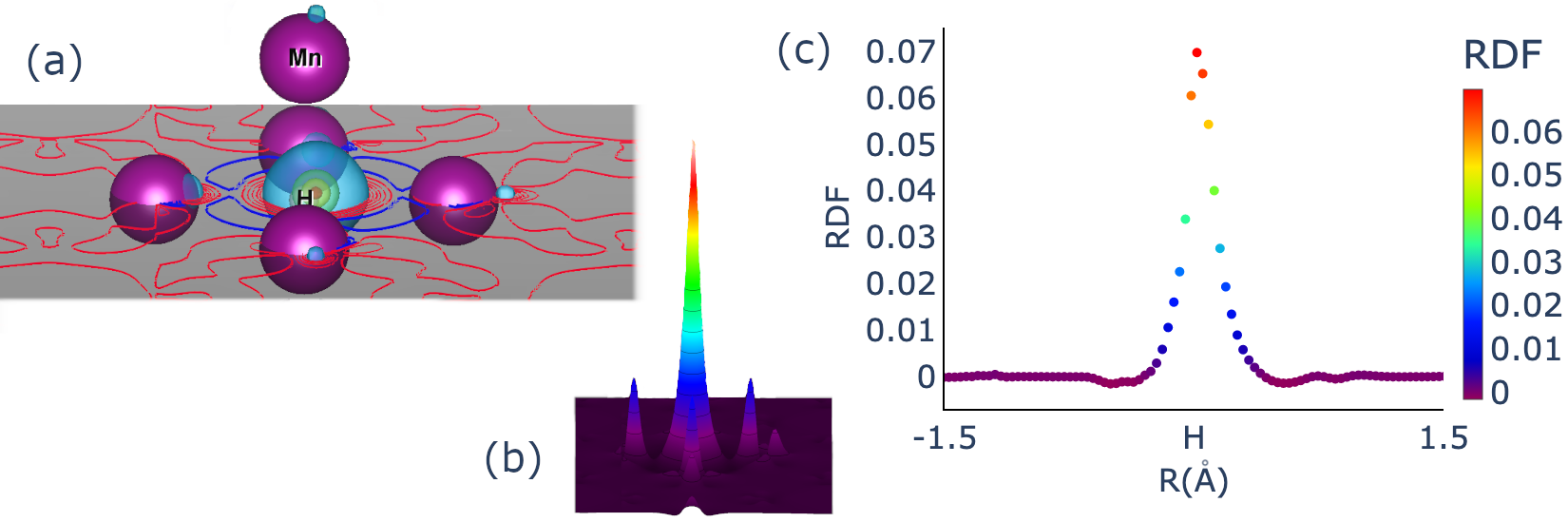}
    \caption{(a) Three dimensional visualization of the PChDP of Mn (OI case), resulting from the subtraction of the differential profiles of the pure crystals from the defected ones; multiple 3d-isolines are highlighted, starting from the highly positive (accumulation) in red ($0.07 \text{ e/bohr}^3$), to positive in yellow and green ($0.05$ and $0.03 \text{ e/bohr}^3$ respectively), to faintly positive, in blue ($0.009 \text{ e/bohr}^3$). On the highlighted gray sectional plane, some accumulation (red) and depletion (blue) isolines are represented, and (b) displays the corresponding bird-eye visualization of the perturbation, with a color gradient for the  charge value mapping the isolines values displayed in the 3D visualization panel, ranging from $-0.0043\text{ e/bohr}^3$ to $0.0759\text{ e/bohr}^3$, and with out-of-plane extension basing on the intensity. (c) The RDF profile obtained from the PChDP.}
    \label{fig:Mn_NotInt_DiffCharge_TOPLOT}
\end{figure*}

\subsubsection{K-Means Clustering}\label{subsec:meth4.2}
Unsupervised learning can be accomplished by combining PCA and K-Means clustering. The K-Means algorithm was proposed by Lloyd in 1982 \cite{kmeans} to solve the problem of finding the minimum total squared distance of each point $x\in X \subset \mathbb{R} ^d$ from its closest center $m \in M$ of cluster $c \in C$. 
\begin{equation}\label{SSD}
    \phi_X = \sum_{x \in X} \underset{m \in M}{min} \norm{x-m}^2
\end{equation}
A survey from ten years ago \cite{kmeanssurvey} reports it already was, and probably still is, the most used clustering algorithm. The problem is solved in the following way:\begin{itemize}
    \item Define arbitrary (tipically random) centers $m_1^{(t=0)}.\,.\,.\,m_k^{(t=0)}$;
    \item Define a cluster $c_i$ as the set of points closer to $m_i$ than $m_j$, for each $j\neq i$;
    \item Define the new k-centers (k-means) as $m_i^{(t+1)}=\dfrac{1}{\abs{c_i}}\sum_{x\in c_i}\,x$
    \item Repeat the previous two steps until the set of means is converged
\end{itemize}
Even though the target total squared distance is a monotonic decreasing function of the number of clusters and, since there are $k^n$ possible clusters, the algorithm is proved to converge, bad clustering can be obtained without a proper initialization. For this reason, Arthur and Vassilvitski introduced a seeding algorithm \cite{kmeans++}, \texttt{K-Means$++$}, able to outperform the results of usual \texttt{K-Means} just by weighting the selected data-points according to their squared distance from the already randomly selected center. \\ A part from a proper initialization, whenever no ground truth is available about the number of cluster in which the data is expected to classify in, there is the need of finding the optimal number of clusters. A variety of methods is available in order to perform the cluster number analysis, and we here focus on the Elbow analysis \cite{elbow}, based on the sum of squared distances~\cite{footnote_elbow} in Eq.\eqref{SSD}: when its decrease with the number of clusters starts to be negligible, the optimal number is found, above which overfitting occurs.



\section{Results} \label{sec:Results}
\subsection{The RDFs}\label{subsec:RDF}
In order to better explain the building and meaning of the OI and TI RDF profiles, let us focus on one specific example, the case of an hydrogen atom as an octahedral interstitial defect in a Mn supercell, shown in Fig.(\ref{fig:Mn_NotInt_DiffCharge_TOPLOT}).

The PChDP isolines can be visualized in the three dimensional structure of the supercell, as in panel (a): the PChDP has its positive maximum (accumulation) close to the hydrogen atom, and fades out to a zero level oscillation with increasing distance from it, as the progressively weaker isolines report. In panel (b) we remove one dimension from the visualization, in order to appreciate the perturbation on a plane that crosses the positive maximum. The color gradient for the charge value is mapping the isolines values displayed in the 3D visualization panel, but is also strictly related to the RDF peak height of panel (c): it leads to the removal of the orienteering dependence and is built as explained in Sec.(\ref{sec:Methods}). 
If one now repeats the extraction for both the OI and TI cases of different samples, as described in Sec.(\ref{subsec:meth2}), obtains the result shown in Fig.S.(7) of the SM. For completeness, some relevant quantities from the plots are reported in Table(III) and Table(IV) of SM, which include peaks heights and widths. 

\begin{figure}[htb]
    \centering
    \includegraphics[width=0.45\textwidth]{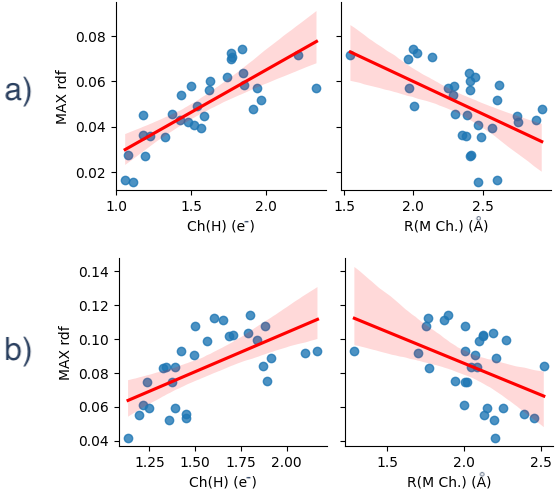}
    \caption{Linear fits and 95\% confidence regression intervals on the relation among RDF peaks properties (MAX rdf) and the charge density ones of the hydrogen (H) and metal atoms (M), both for OI (a) and TI (b) configurations.}
    \label{fig:LinearRelations}
\end{figure}

It is important to notice how even at this level, without the support of any ML model, one is able to make some compositional-based comparison between the investigated cases, by looking at the peaks properties.
An intuitive but strong relation providing a first insight on the presented curves could be found in the charge densities themselves. If one considers the obtained maxima values for the RDF profiles and the properties of the charge density profiles coming from Bader charge analysis, as in Fig.(\ref{fig:LinearRelations}), good linear relations are found, especially with the hydrogen Bader charges. The finding suggests the tendency of having higher RDF peaks for larger accumulated charges on hydrogen, which stands as reasonable thinking about the RDF perturbation as a representative of the charge density response (screening) to the added impurity, and given that the RDF profiles themselves are built on the hydrogen PChDPs. However, a clearer picture requires the exploration of correlations with the entire spectrum of atomic-structural and density properties, as will be proposed in  Sec.(\ref{subsec:correlations}). Here we would like to stress that the RDF itself is providing an isotropic measure of the PChDP in the crystal due to the insertion of the H atom, and its comparison for different metallic crystals can clearly provide a measure of similarity of the charge response. For this reason, it can be considered as the descriptor of the perturbed profiles, being a unique signal and fingerprint of the defect-crystal interaction.

In the next paragraph, we question the possibility of a unique average descriptor, on the basis of which a first classification of the system-specific defect-induced perturbations can be demonstrated. 

\subsection{Composition specific fluctuations from an average oscillation} \label{subsec:universal}

\begin{figure}[htb]
    \centering
    \includegraphics[width=0.44\textwidth]{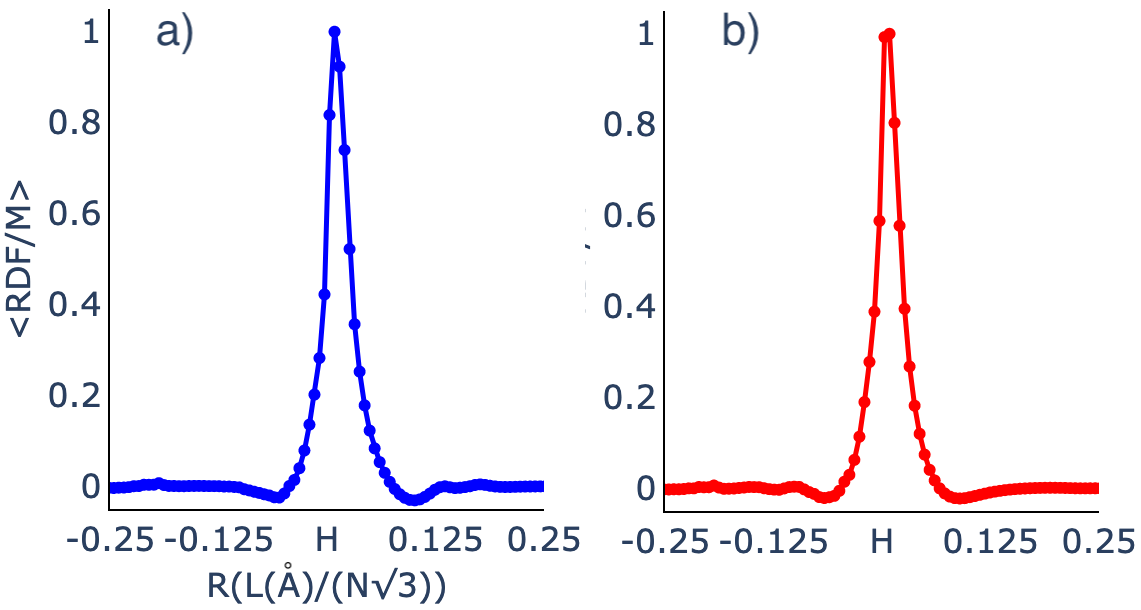}
    \caption{RDFs averaged over all the available configurations (Eq.\eqref{eqn:mfosc}), for both OI (a) and TI (b) states. The axes share the same labels.}
    \label{fig:UnivCur}
\end{figure}

 It is of interest the possibility of considering a H-\textit{universal} RDF curve for our problem, capable of grasping the average behaviour of all the system specific charge density responses that we have observed, assuming crystal composition as a degree of freedom. In order to do so, one could consider the average over $N$ different bulk crystals
\begin{equation} \label{eqn:mfosc}
    \langle \dfrac{\text{RDF}_i}{M_i} \rangle\rvert_{i=1..N} \,\,,\,\, M_i = max (\text{RDF}_i)
\end{equation} which is proposed in Fig.(\ref{fig:UnivCur}). The resulting averaged curves of OI and TI cases are very similar, a part from an inverted symmetry noticeable in the short-range rippling of the profiles.

\begin{figure*}[htb]
    \centering
    \includegraphics[width=0.99\textwidth]{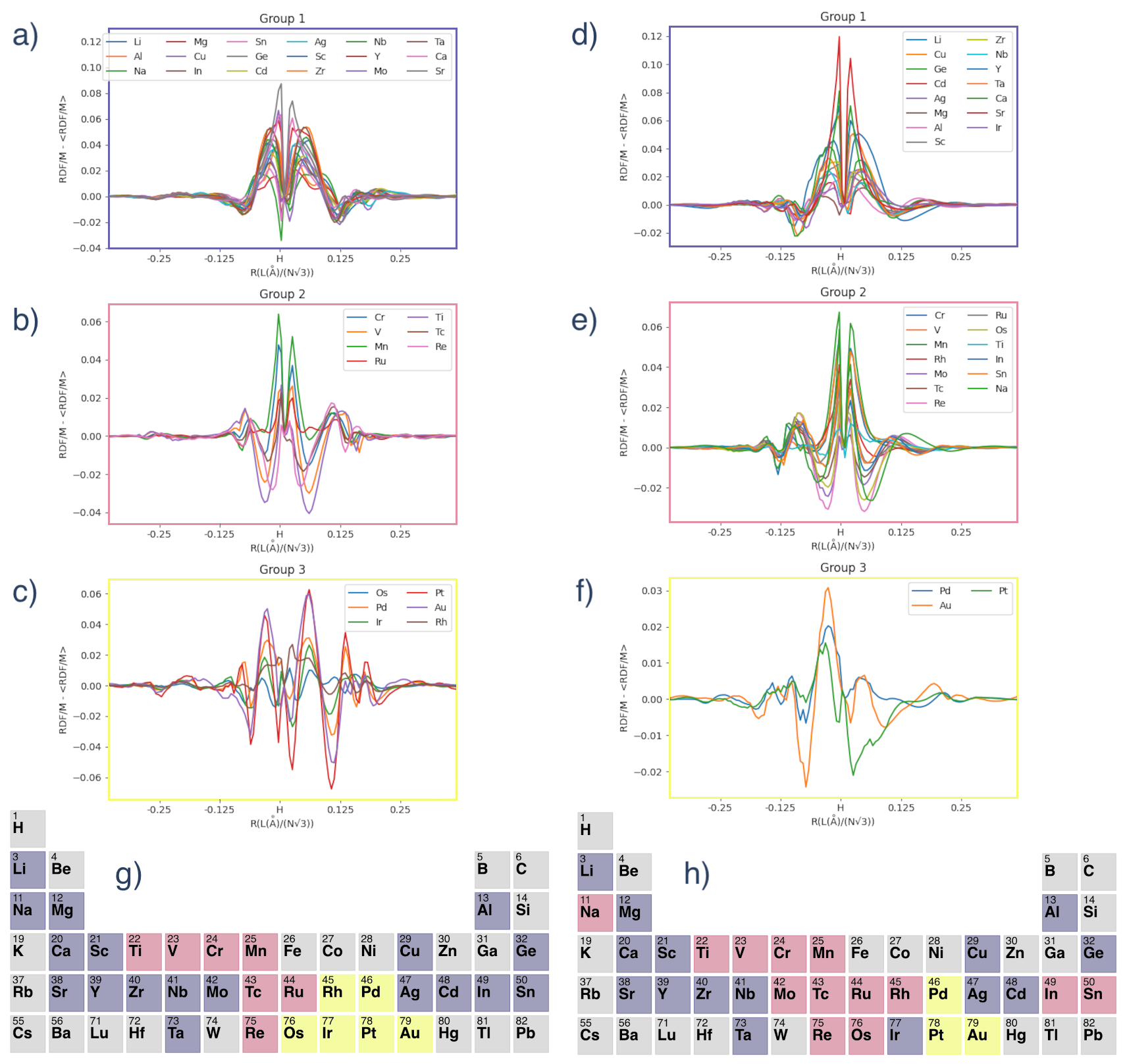}
    \caption{Crystal-specific oscillations from the rescaled and averaged RDF (Eq.\eqref{eqn:deviations}), both for OI (a: Group1, b: Group2, c: Group3) and TI (d: Group1, e: Group2, f: Group3) states. The three behavioural groups are highlighted by a colored plot frame, which helps highlighting them in the periodic tables below (g: OI, h: TI). }
    \label{fig:Fluctuations}
\end{figure*}

In order to be able now to quantify the capabilities of a coarse-grained perturbation of grasping the peculiarities of the system specific ones, let us consider the deviations from the average
\begin{equation} \label{eqn:deviations}
     \dfrac{\text{RDF}_i}{M_i} \,\, - \,\, \langle \dfrac{\text{RDF}_i}{M_i} \rangle\rvert_{i=1..N} \,\, , \,\, M_i = max (\text{RDF}_i)
\end{equation}
which are reported in Fig.(\ref{fig:Fluctuations}), both for OI and TI states. 

Interestingly, again without any ML support, the deviations can be easily grouped in three behavioural classes basing on the minima and maxima positions, which also show a conservation in their overall characteristics going from the OI to TI state, a part, as for the averaged RDFs, from an inverted symmetry along the axis of distances from their center (hydrogen).  As it can be noticed from the periodic table plots, the majority of the group members are conserved going from the OI to TI state, except for few changes involving Na, Mo, Rh, In, Sn, Os and Ir.

 Going from Group 1 to Group 2, which together include almost all the non-noble metals, the local deviations from the average charge perturbation gain two global minima which are almost symmetric with respect to their center: thinking in terms of the original characteristics of the signals appearing in the operation of Eq.\eqref{eqn:mfosc}, this should imply a change in the width of the crystal specific one, with respect to the averaged perturbation. Interestingly, we notice how noble-like metals, namely Au, Pd, Pt, Rh and Ir, are classified inside Group 3 (OI), and are characterized by a considerable increase in the number of maxima and minima of the presented deviations. This special class of metals is generally known for their stability, resistance to corrosion and oxidation in moist air, as well as being effective catalysts for the hydrogen evolution reaction (HER)~\cite{noblemetals}. Other transition metals from Group 2, such as Ti, V and Cr, when combined to form an high-entropy alloy (HEA), have been proved to absorb hydrogen at room temperature and with moderate equilibrium pressures for the dihydride formation~\cite{HEAs}.

Even though verifying the actual catalytic activity or hydrogen storage performances of the proposed crystals, or their combinations, in the presence of hydrogen molecules is beyond the scope of the present work, we believe these observations inspire ourselves, and possibly the readers, for further work in this direction. It is worth to point out that the proposed element-specific deviations are a sophisticated form of characterization of the metal-hydrogen interaction, inheriting the richness of the presented descriptors, and delivering a bias-free classification of the local peculiarities of the charge density responses due to characteristic atomic (therefore, electronic) environments. One may argue the inter-classes variations in the number of extremas to be related to a change in the energy of the electrons providing the screening of the impurity: a longer rippling profile shall be associated to lower energies, and indeed, as we will show in the next sections, the accumulated charge on H is significantly minimized in the noble metals.

In the next paragraph, the capabilities of our descriptors to correlate with the physical properties of the systems are quantitatively investigated, and will shed light over the intertwined relations driving the presented classification. In Sec.(\ref{subsec:densitydataset}), an alternative and more traditional classification method will be presented basing on a set of density-related features, investigating their actual influence on the samples distribution.

\subsection{The density dataset: correlations}\label{subsec:correlations}
\begin{figure}[H]
    \centering
    \includegraphics[width=0.45\textwidth]{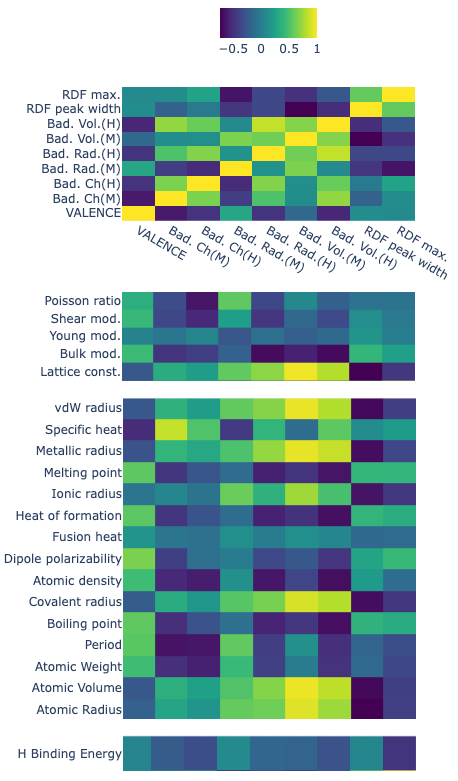}
    \caption{Correlation matrix between the density-related features (x-axis) with themselves (higher y-block), with the structural (second y-block) and atomic (third y-block) features, and with the computed hydrogen binding energy (lower y-block).}
    \label{fig:Dens-Corr}
\end{figure}

In order to prove the key role of the PChDP as effective descriptors, a dataset has been built on the base of the density information, from the Bader charge and RDF analysis, as well as on the structural and atomic one, collected for each metallic crystal here considered. The Pearson correlation matrix for the dataset can guide our attention on the most relevant features, and, given that the relationships between atomic and structural properties are well known, in Fig.(\ref{fig:Dens-Corr}) we focus on some specific sub-blocks. 

\begin{figure*}[htb]
    \centering
    \includegraphics[width=0.85\textwidth]{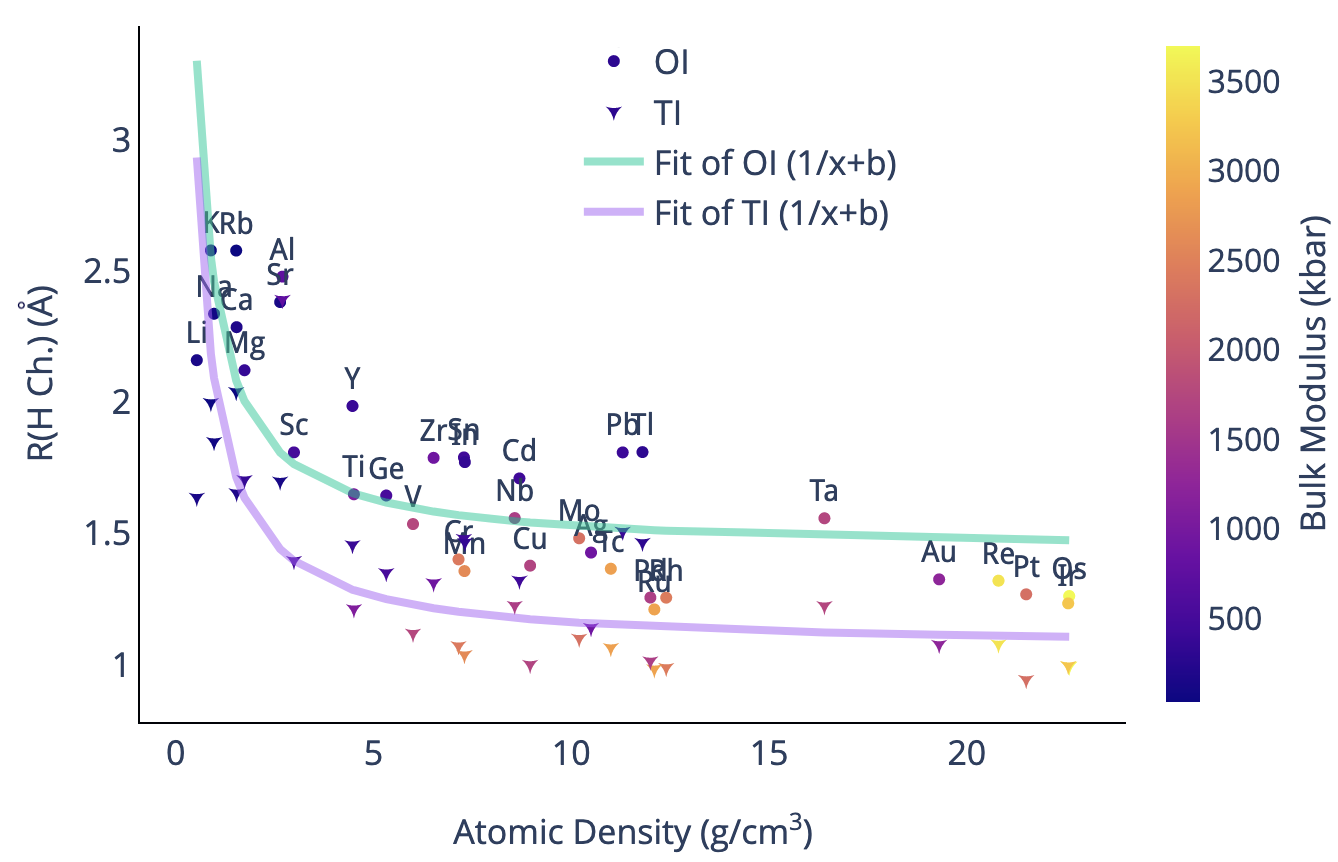}
    \caption{Plots of specific relationships involving atomic (atomic density), structural (bulk modulus) and electron density (H Bader radius) features for both OI and TI samples, with the superimposed visualization of fitted trend-lines.}
    \label{fig:Expl_Correlations_2}
\end{figure*}

Let us focus first on the density related features. Some expected behaviour is obtained: i) M bader charge and H bader charge exchanges show a remarkable positive correlation (0.7 in scale), implying the exchange from M to H (Tab.S.(III,IV) in S.M.) ii) the larger is the number of valence electrons of the metals (M), the smaller the charge it exchanges with (to) hydrogen (H); iii) the two definitions of bader charge extension in terms of radius and volume are largely correlated; iv) low correlation is present between valence and radius of the M charges, due the expected correlation of the latter with the definition of the atomic radius~\cite{footnote_correlation}, and therefore the periodic behaviour along the periodic table.
Moreover, the maximum value and the width of the computed RDF peaks in the grid of the hydrogen (H) PChDP is strongly anti-correlated with the Bader charge extension of the metal (M) both in terms of radius~\cite{footnote_radius_extensions} (-0.6 and -0.5 respectively in scale) and volume (-0.5 and -0.7 respectively in scale), and hydrogen charges properties are affecting the RDF peaks in an overall similar way. 
From the similar behaviour of M and H charge extensions, the similar influence of the latter on the RDF peaks properties can now be understood. 

Let us now consider the correlations between the density related features and the atomic features of the involved metals. Interestingly, the RDF peaks widths show strong anti-correlations with the metal atomic radius, atomic volume and any other definition of bonding radius, and the reason lies in the close relationship that the RDF peaks have with the metal and hydrogen charge features: the bader charge extensions of the metals are fully correlated with the definitions of atomic volume and radius, and very tight relationships also hold with hydrogen charge extensions. 

Coming back to Fig.(\ref{fig:Fluctuations}), the reader can now understand the sensitivity of the descriptors properties to changes in the host crystalline matrix composition, which is at the base of the previously emerged classification: for example, discrepancies in widths between crystal-specific and a-specific perturbations (averaged) drive the oscillatory behaviour of their difference, and give meaning both to an average screening, as the result of average atomic properties, as well to interstitial hydrogen as a probe of the composition-specific properties, through the induced electronic perturbation. It is also interesting to notice how non-negligible correlations are also found with larger-scale properties like melting points, heat of formation, boiling point. 

Concerning the correlations between the density related features and the structural ones~\cite{footnote_structuralprops}, again the PChDP information grasps very efficiently the properties of the host crystals: the Poisson ratio and shear modulus are both encoded in the assumed hydrogen bader charge with a strong anti-correlation (-0.6 in scale), while the RDF peaks properties are closely related to the lattice constants. Visualizations of the discussed intertwined relations of atomic and structural properties with the density descriptors are shown in the example of Fig.(\ref{fig:Expl_Correlations_2}). 

The remarkable correlations that Fig.(\ref{fig:Dens-Corr}) highlights among the hydrogen BEs and charge related properties, like the RDFs maxima or the hydrogen Bader chargers, have led the focus over this quantity and its variations among the different crystals, and corroborates the thesis of effectiveness of the proposed electron density fields as descriptors of the system interactions. 
\begin{figure*}[htb]
    \centering
    \includegraphics[width=0.95\textwidth]{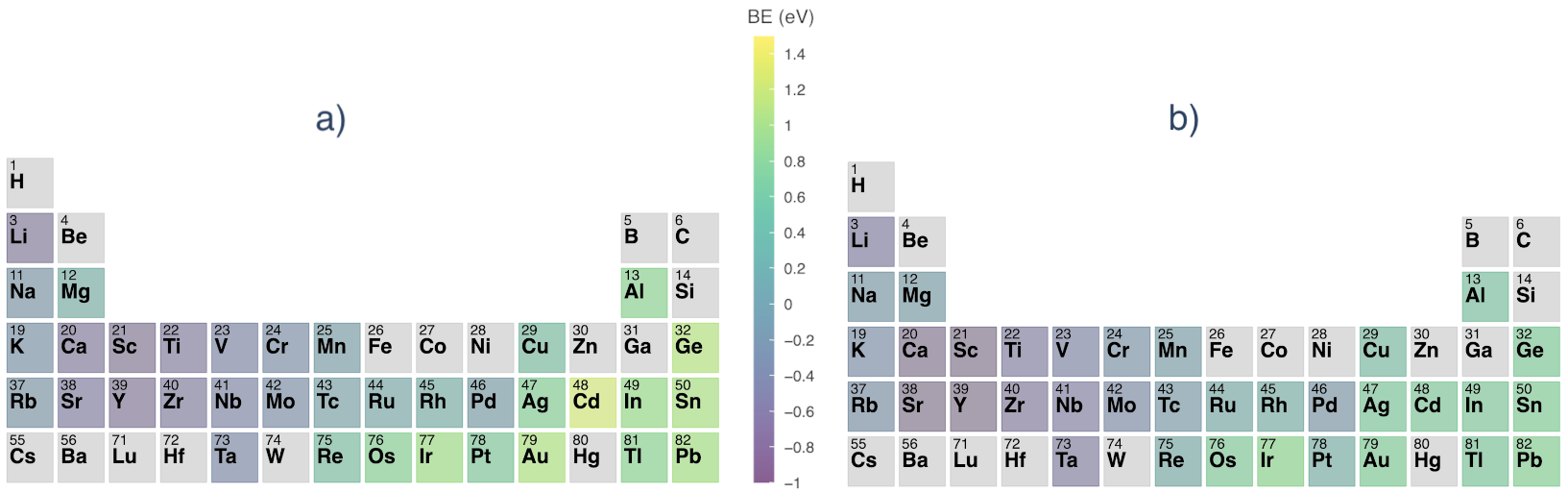}
    \caption{Periodic table plots of the hydrogen binding energy (BE) in the OI (a) and TI (b) states of the considered FCC crystals. The colorscales are unified to the one of the single colorbar displayed. The bulk crystals that have not been considered in this study have been either removed (lanthanoids, actinoids, groups 15-18, period 7) or left in grey. }
    \label{fig:BEptable}
\end{figure*}

Fig.(\ref{fig:BEptable}) reports the computed BE values on periodic tables for both OI and TI states. Overall, a part from few outliers, the binding tends to gradually increase along the periods, with negative values among the alkali, alkaline earth and early transition-metals, where it crosses the zero level to become positive among the late transition-metals, increasing up to the crystallogens. It is to notice that the possibility of obtaining composition maps that could drive the design of systems basing on key properties like the hydrogen binding energy is highly desirable in the context of materials for hydrogen storage: the U.S. D.O.E. has fixed indeed a specific range of interest for this value \cite{SATYAPAL2007246} in hydrogen applications. However, a reliable descriptor is needed to capture the composition induced variations, basing on its underlying power to encode local environments and interactions. In Fig.S.(8) of SM, we propose a visualization of the relation between the hydrogen Bader charges and its binding energy, which is found to be almost linear.

As a last point of this discussion, we want to prove that charge density can help in directly accessing also to the diffusion properties of the hydrogen atom in the different crystals, by looking for a relationship between its migration barriers and charge spatial extensions. In order to prove it, following the procedure explained in Sec.(\ref{subsub:meth3.1}), Nudge-Elastic-Band calculations have been applied for two cases: Al (high volume of hydrogen Bader charge), Mn (low volume of hydrogen Bader charge). 
\begin{figure}[H]
    \centering
    \includegraphics[width=0.47\textwidth]{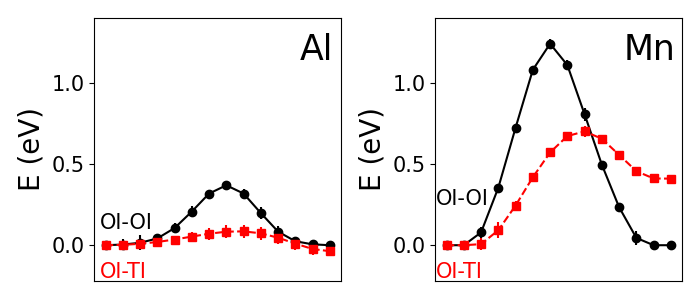}
    \caption{Hydrogen migration energy barriers in Al (more extended Bader charge volumes) and in Mn (less extended). The barriers are the result of climbing-image NEB calculations, as explained in Sec.(\ref{subsub:meth3.1}).}
    \label{fig:NEBplots}
\end{figure}
\begin{table}[htb]
    \centering
    \begin{tabular}{c|c|c}\hline
    &OI-OI E(eV)&OI-TI E(eV)\\ \hline
    Al&0.369&0.085 , 0.122 \\
    Mn&1.241&0.701 , 0.291 \\ \hline
    \end{tabular} 
    \caption{Results from the Nudge-Elastic-Band analysis of the energy barrier for the hydrogen migration along a path in different crystals. For the OI-TI path, two values are present because the barrier evaluation is different depending on whether is computed relatively to the OI (first value) or TI (second value) extremas of the path.}
    \label{tab:NEB}
\end{table}
The results in Fig.(\ref{fig:NEBplots}) and Tab.(\ref{tab:NEB}) show a much smaller energy barrier for the high charge volume case (in Al) with respect to the low charge volume one (in Mn). The full migration energy barrier profiles along the proposed paths also show a similar behaviour in the transition states: the favourite path is OI-TI, as also highlighted in other works for the case of Al\cite{Kaxiras_Al}. The results also lead to an interesting comparison with a specific result in the literature, which, to our opinion, makes both findings more complete: L. Messina \textit{et al.} \cite{Luca} have taken into account a complementary approach for the investigation of the diffusion of impurities in bulk FCC iron; instead of systematically changing the species forming the pure FCC bulk while keeping the same impurity, as done in the present work, they have been keeping the same crystal, systematically varying the interstitial impurity. Even if hydrogen was not involved, among their findings is a counter-intuitive but largely justified correlation between the migration barriers of a solute in iron and the solute compressibility: the larger solutes diffuse faster because of lower migration barriers (a part from P and Si, in their study); moreover, the migration barriers (diffusion coefficients) are displaying a parabolic trend for 4d and 5d elements and an M-shaped (W-shaped) trend for the 3d ones. \\ In the present problem of systematically changing the hosting matrix around interstitial hydrogen, the extracted density information (i.e. its Bader charge volume) is closely related to the mobility of the interstitial in the system and to the compressibility of the latter, in its pure conditions: the larger volumes (smaller transition barriers) are found for larger compressibilities of the pure hosting crystals, in systems where the number of valence electrons of the metals tend to be smaller.

After having assessed the strong descriptive power of the electron density features in grasping a wide spectrum of properties of the systems of interest, in the next section we restrict to the former features and apply unsupervised ML techniques to test their ability to uncover hidden patterns and behavioural classes.


\subsection{Unsupervised Machine Learning on a pure density dataset}\label{subsec:densitydataset}
Let us consider the application of unsupervised ML techniques to the previously presented dataset, in order to possibly reveal hidden patterns in the data that can lead to deeper understanding and characterization. At this stage, we report the result of the analysis on the pure density features dataset. As explained in Sec.(\ref{subsec:meth4.1}), PCA is a powerful procedure to reduce the dimensionality of the dataset down to its most relevant components. 
The histogram in Fig.S.(2) of SM provides the explained variance ratios by each of the resulting principal components: the first two, are able to account for around $70\%$ of the total variance, but, being the result of a transformation of the original dataset components, the individual role of the original features is not clear at this stage. 

\begin{figure}[htb]
    \centering
    \includegraphics[width=0.4\textwidth]{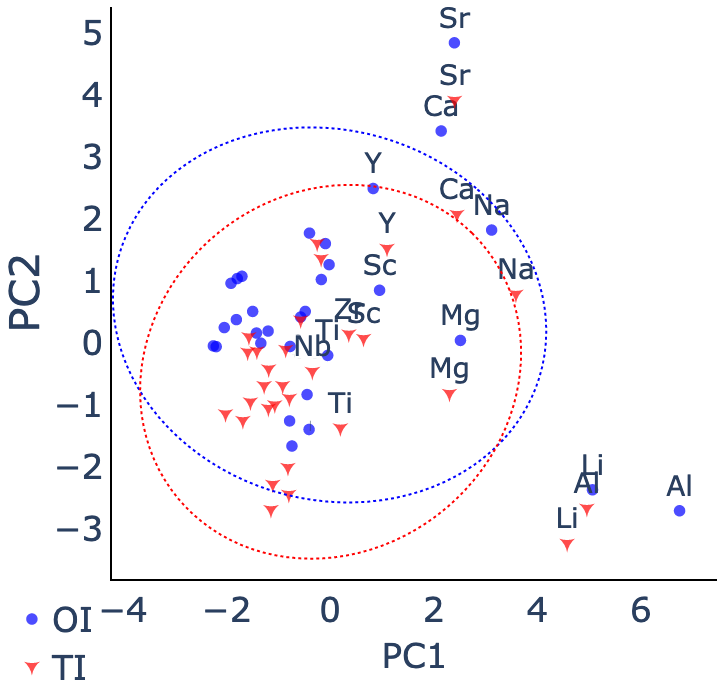}
    \caption{Visualization of OI and TI samples in the first two-PCA components space, together with the corresponding $95\%$ confidence ellipses. Labelling has been limited to samples out of the crowded region, in order to ease the reading of the plot.}
    \label{fig:VisualClust}
\end{figure}

In order to try to understand whether an overall classification could emerge in the form of behavioural classes among the investigated samples, clustering methods can be applied. Since we know that the samples are characterized by two different hydrogen interstitial positions in the bulk host crystals, namely OI and TI, the most naive way of tentative clustering could be to highlight the two in the reduced PCA space and evaluate their separation, as shown in Fig.(\ref{fig:VisualClust}). As it can be noticed, the two clusters have a similar distribution in the first two principal components space, but slightly separated. The reason for the shifting lies in the electron density information, the only one differentiating the OI and TI samples at this stage. However, resorting to separation in intuitive classes does not necessarily help with the emergence of meaningful local clusters in our dataset, or with the understanding of the role of each specific electron density feature in driving the samples distributions. 
In absence of prior knowledge on the number of classes that our dataset could distribute among, the K-Means method, introduced in Sec.(\ref{subsec:meth4.2}), is among the ones that can be adopted to perform cluster analysis on the reduced space of the principal components. A detailed evaluation of the optimal number of clusters should be performed in the absence of ground truth, and, as shown in Fig.S.(3) of SM, the Elbow method indicates $k=5$ as the number of clusters to optimally group our data.

A quantification of the influence of each feature is necessary and can be accessed through the so called \textit{biplots}, in which the loadings are displayed in the reduced spaces of the principal components. As discussed in Sec.(\ref{subsec:meth4.1}), the loadings represent the covariances between the PCs and the original variables and therefore are informative on the variables variation explained by the components.
\begin{figure*}
    \centering
    \includegraphics[width=0.99\textwidth]{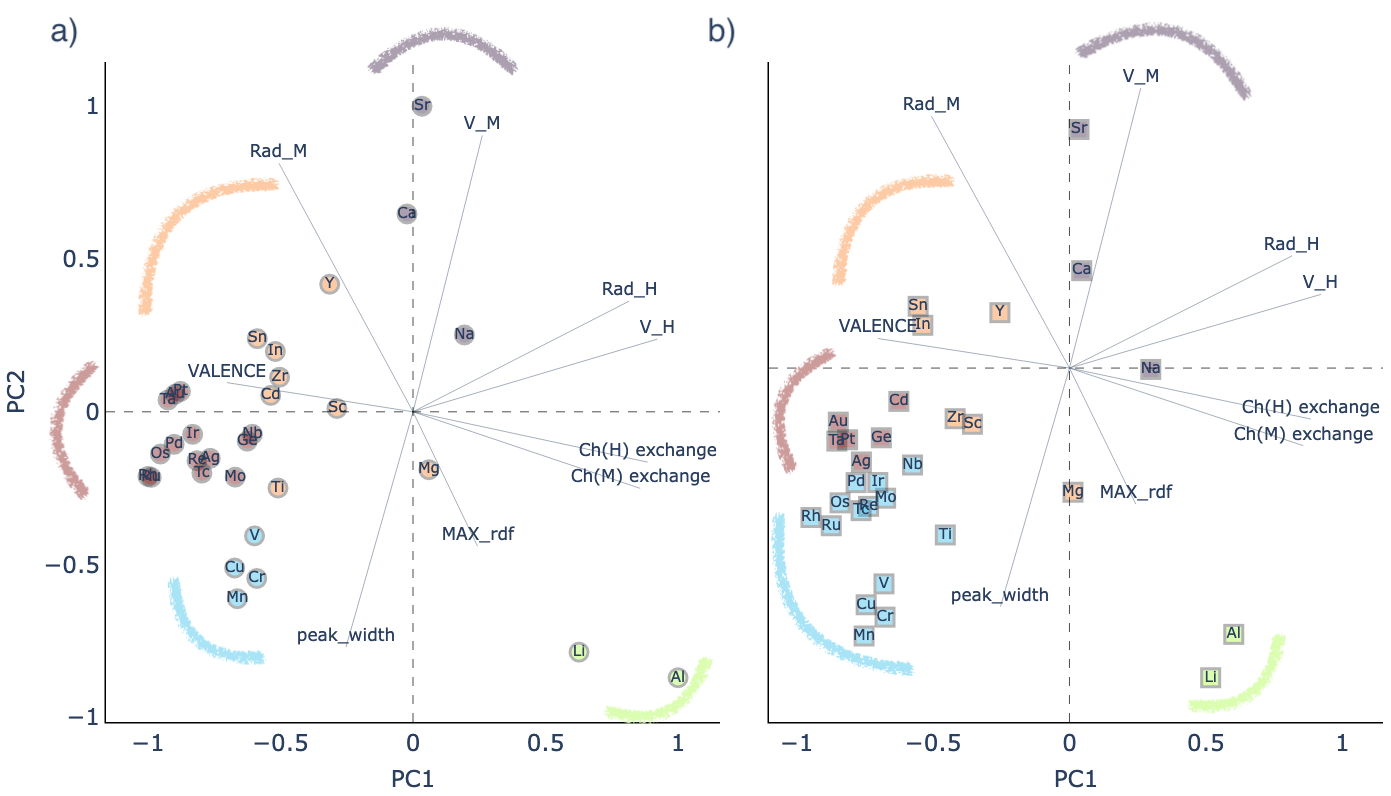}
    \caption{The biplot for the first and second principal components, both in OI (a) and TI (b) configurations. The plot is a representation of the loadings on the two-components subspace, allowing to recognize the influence of each feature on the components, and thus their driving on the clusters distribution. In order to allow for a comprehensive visualization of the biplot, the PCs are rescaled to the $[-1,1]$ interval. The colours for the scattered dots come from the K-Means clustering labels for the optimal number of clusters, and colored arcs highlight the plane sectors where the corresponding clusters are found. The loadings values are presented in the Supplementary Material in Tab.S.(VI).}
    \label{fig:Loadings1}
\end{figure*}
In the biplots of Fig.(\ref{fig:Loadings1}), the clusters have been highlighted following the optimal K-Means clustering previously analysed, in order to be able to associate the influence of the original variables also on the clusters distribution, besides the overall one. As it can be noticed, out of the optimal clusters, some are noticeably smaller: the one of Al and Li, i.e. maximizing the charge exchange from M (in terms of a loss) to H (in terms of a gain), or the one of Sr, Ca and Na, maximizing the measures of the metal charge extensions, minimizing at the same time the RDF peak properties. Noticeably, transition metals form three close clusters, all in favour of minimizing the amount of charge exchanged to hydrogen, and so its charge extensions. As seen in the previous subsection, charge extensions are related to migration energy barriers between high-symmetry interstitial states, and this analysis represents an example of direct access to configurational and compositional subspaces exploration for tuning the defect influence on the electron density local landscape, which, as shown, grasps a large variety of physical properties of the host matrix, and of the defect-host interaction, as we believe a general effective descriptor should do.


\section{Final Remarks} \label{sec:FinalRemarks}
Through a detailed correlation analysis involving the properties of the ab-initio defect-induced charge density perturbations, we prove the ability of the latter to effectively grasp atomic and structural properties of their host metal systems. The expected strong correlations between the induced charge perturbation profiles and the underlying captured atomic charge properties, suggests the role of the latter in the long-bridging abilities of the former with a wide spectrum of crystal properties. The power of the proposed approach lies in its simplicity and generality: any defecting or functionalization of a system introduces a perturbation in the charge density profile, which, embodying the guest-host interaction, naturally inherits features of both on a pure density basis. Further reasoning comes from the brief analysis of the correlation between the hydrogen charge extensions and its transition energy barriers, for candidate systems in which the former quantity presented large differences: where hydrogen displays larger charge extensions, there are found smaller energy barriers for transitioning along different paths. This interesting result shall be further investigated, but clearly defines a strategy of predicting defects mobility just on a density-based argument. The power of the density related features in driving the distribution of the samples in a classification task has been addressed, in turn highlighting the importance of the density-aided classification task itself in studies of defects. The resulting maps, containing behavioral classes spanning different compositions and configurations, can effectively drive material searches basing on maximization/minimization of density properties, knowing their quantified strong influence on the "phenotypical" atomic/structural properties, being the density the well known associated "genotype", in this biological analogue. A detailed analysis of crystal-specific charge density response variations from a compositionally coarse-grained one has also been performed, leading to the spontaneous emergence of behavioural groups showing the same oscillatory peculiarities. Further investigation is needed in order to fully correlate the gradual changes in the charge responses of the different crystals with complex phenomena like catalysis or storage performances. However, the power of the approach and of the new proposed density-based descriptors gives hope and inspiration for deeper and larger scale investigations along these important directions. 
\section*{Acknowledgements} \label{sec:acknowledgements}
We acknowledge support from the European Union Horizon 2020 research and innovation program under grant agreement no. 857470 and from the  European Regional Development Fund via the Foundation for Polish Science International Research Agenda PLUS program grant No. MAB PLUS/2018/8.

\bibliography{bibliography}%

\begin{thebibliography}{77}
\bibitem [{foo({\natexlab{c}})}]{footnote_elbow}%
  \BibitemOpen
  \href@noop {} {} ({\natexlab{c}}),\ \bibinfo {note} {it is to notice how this
  quantity is related to compactness of the clusters.}\BibitemShut {Stop}%
\bibitem [{foo({\natexlab{d}})}]{footnote_correlation}%
  \BibitemOpen
  \href@noop {} {} ({\natexlab{d}}),\ \bibinfo {note} {it will be shown
  explicitly soon in the same section}\BibitemShut {NoStop}%
\bibitem [{foo({\natexlab{e}})}]{footnote_radius_extensions}%
  \BibitemOpen
  \href@noop {} {} ({\natexlab{e}}),\ \bibinfo {note} {the radius of the M
  charge here behaves differently from the volume, and it may be due to its
  definition in the bader extraction program, as the minimum distance from the
  Bader surface. In this view, volume can be seen as a more reliable estimate
  of charge extensions due to anisotropies and bondings. We believe this to be
  reason also for the very high correlation between V(M) and V(H), and less
  pronounced one between R(M) and R(H).}\BibitemShut {Stop}%
\bibitem [{foo({\natexlab{f}})}]{footnote_structuralprops}%
  \BibitemOpen
  \href@noop {} {} ({\natexlab{f}}),\ \bibinfo {note} {these are properties of
  the metal bulk crystals before being defected with interstitial
  hydrogen.}\BibitemShut {Stop}%
\end{thebibliography}


\end{document}


\title{Supplemental materials to the manuscript}
\maketitle




\section{Size effects}
\begin{figure}
    \centering
    \includegraphics[width=0.9\textwidth]{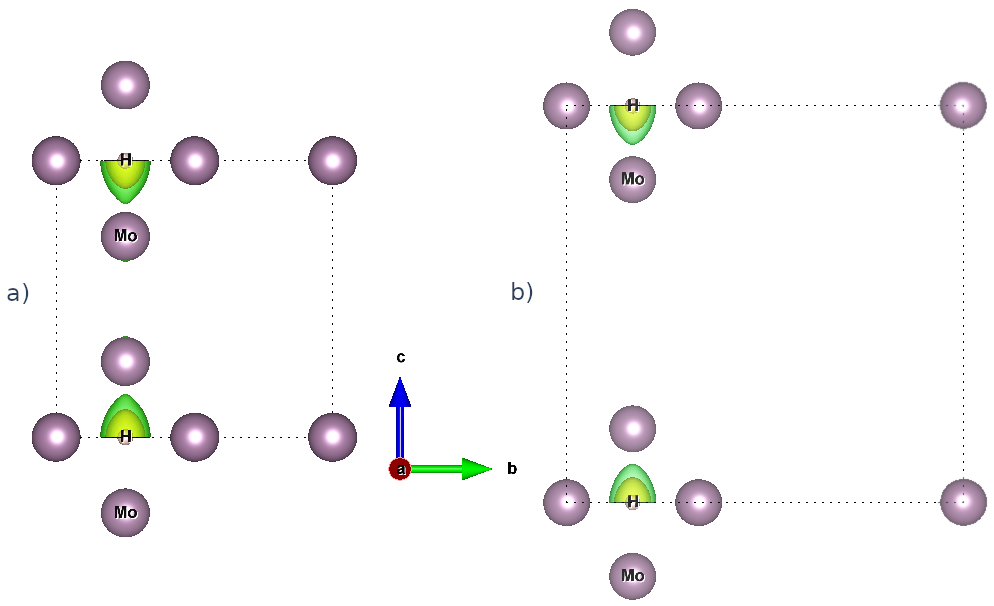}
    \caption{Visualization of the three dimensional differential electron charge perturbation induced by an hydrogen interstitial defect in a $2\times2\times2$ (a) and $3\times3\times3$ (b) supercell of Mo. Two same isoline levels, in green and yellow, are visible in the plots, and their values are reported in Tab.S.(V), together with the minimum and maximum isolines values of the two grids. For simplicity of visualization, just the H-neighbouring Mo atoms and their periodic replicas are reported in the FCC supercells.}
    \label{fig:sizeffects}
\end{figure}

In Fig.S.(\ref{fig:sizeffects}) we propose a comparison between the hydrogen-induced electron charge density perturbation isolines in $2\times2\times$ and $3\times\times3$ supercells, considering Mo as an example. As shown in the main results of the paper, the Radial Distribution Functions in the density grids are dominated by positive (accumulation) peaks, coming from a dominance of charge accumulation, rather than deplition, in the different matrices. Fig.S.(\ref{fig:sizeffects}) proves high compatibility among the two supercell sizes results, therefore we believe that a studied carried on $2\times2\times2$ supercells would not suffer from relevant size artifacts. 

\section{Unsupervised learning}
\begin{figure}[htb]
    \centering
    \includegraphics[width=0.5\textwidth]{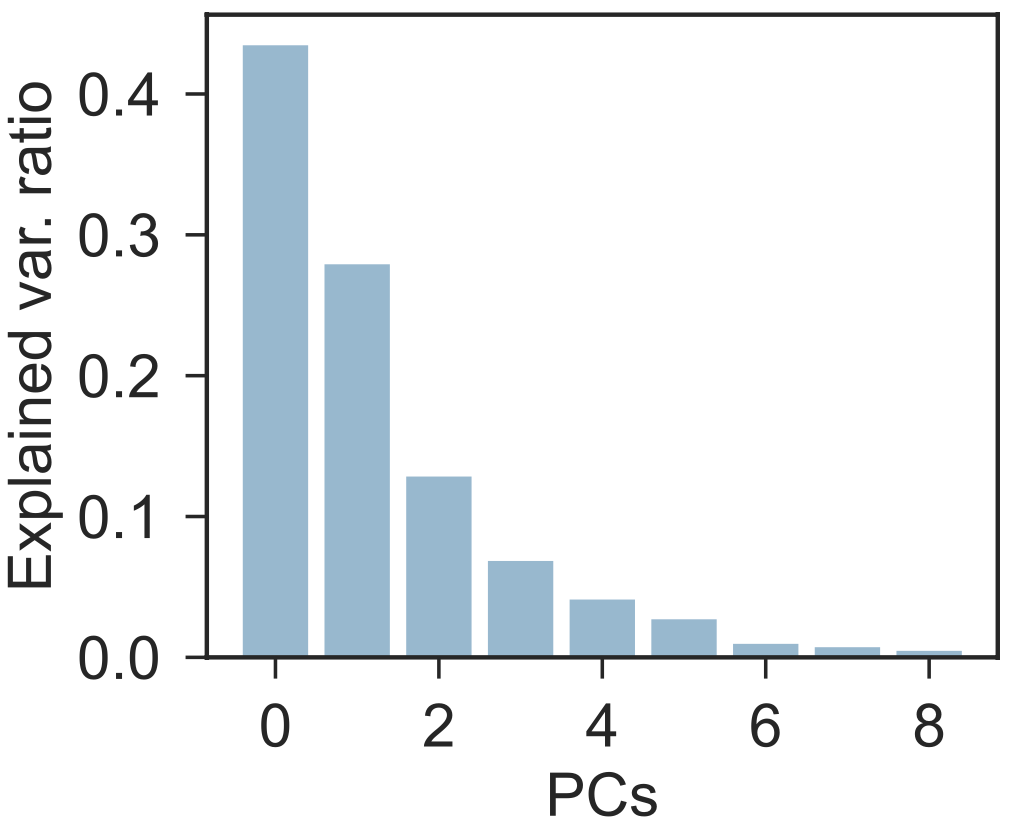}
    \caption{Histogram of explained variance ratios by each of the resulting principal components after applying PCA: the first two, are able to account for about $70\%$ of the total variance.}
    \label{fig:PCAcomps_ex_var}
\end{figure}

\begin{figure}[htb][htb]
    \centering
    \includegraphics[width=0.4\textwidth]{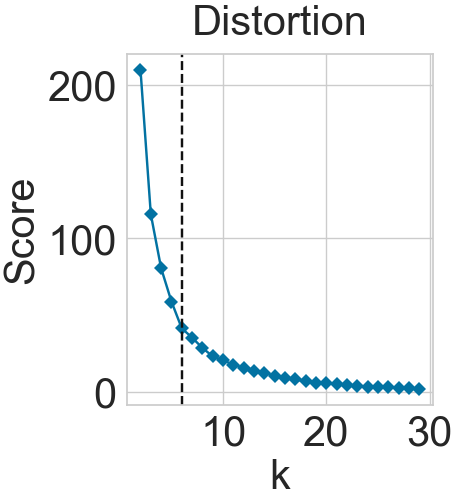}
    \caption{The Elbow method score as a function of the number $k$ of K-Means clusters. The test indicates $k=5$ as the number of clusters to optimally group our data.}
    \label{fig:elbow}
\end{figure}

In the paper to which this material is supplementary, it has been shown the use of density information coming from different crystals in dataset to be used for unsupervised learning tasks. Among these tasks, a standard technique is Principal Component Analysis, and the evaluation of the explained variance ratio captured by each resulting component, visible in Fig.S.(\ref{fig:PCAcomps_ex_var}). As visible from the histogram, the first two principal components are able to account for more than 70\% of the total variance. In the space of the two main principal components, clustering methods can be applied to highlight the emergence of behavioural classes, and we have decided to proceed with K-Means. In absence of ground-truth on the number of clusters, the Elbow method can be applied to find out the optimal one, as in Fig.S.(\ref{fig:elbow}). 

\section{Different symmetries}

\begin{figure}[htb]
    \centering
    \includegraphics[width=1.0\textwidth]{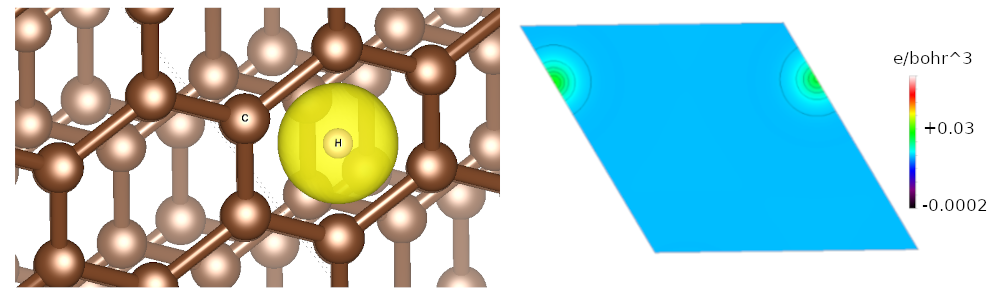}
    \caption{Three dimensional (left) and two dimensional visualization (right) of the differential electron charge density perturbation induced by the hydrogen defect in a HCP C $2\times2\times2$ supercell. The isoline value on the left is $0.0008 \text{ e/bohr}^3$. }
    \label{fig:C}
\end{figure}

\begin{figure}[htb]
    \centering
    \includegraphics[width=1.0\textwidth]{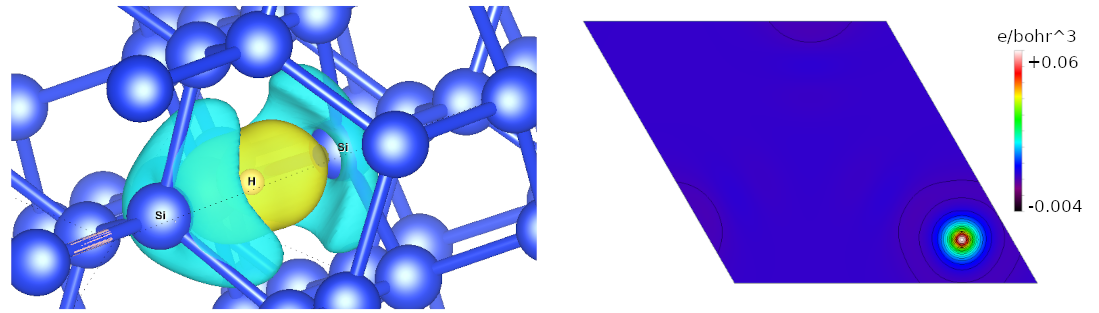}
    \caption{Three dimensional (left) and two dimensional visualization (right) of the differential electron charge density perturbation induced by the hydrogen defect in a HCP Si $2\times2\times2$ supercell. The isoline value on the left is $0.0020 \text{ e/bohr}^3$.}
    \label{fig:Si}
\end{figure}

\begin{figure}[htb][htb]
    \centering
    \includegraphics[width=1.0\textwidth]{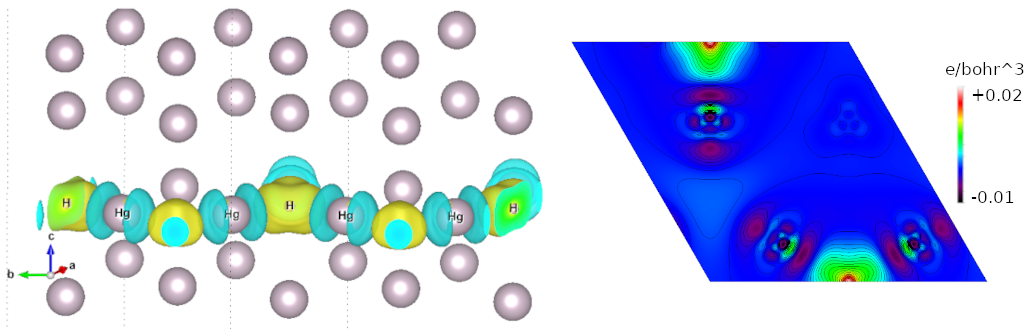}
    \caption{Three dimensional (left) and two dimensional visualization (right) of the differential electron charge density perturbation induced by the hydrogen defect in a HCP Hg $2\times2\times2$ supercell. The isoline value on the left is $0.0010 \text{ e/bohr}^3$.}
    \label{fig:Hg}
\end{figure}

In the main text of the work, we have presented results regarding hydrogen-induced electron charge density perturbations in a variety of FCC metallic bulk systems. We found interesting the idea of providing few examples of the same densities in non-metallic bulks of different symmetries, like HCP. In Fig.S.(\ref{fig:C},\ref{fig:Si},\ref{fig:Hg}) we provide these examples for one interstitial hydrogen atom living in $2\times2\times2$ supercells of C, Si and Hg, respectively. Intentionally, we selected interstitial sites without paying attention to them being highly symmetric, in order to explore the characterization of the perturbations in various conditions. In HCP C, the hydrogen atom sits in between two layers of the bulk structure. The pertubation symmetry is locally spherical and resembles the ones examined in FCC crystals. In HCP Si, the perturbation is still predominantly characterized by charge accumulation with slight elongated distorsion towards the Si atoms highlighting bonding. In HCP Hg, hydrogen-induced charge depletion gains importance and reaches the same order of magnitude as accumulation. In this latter case, the hydrogen fingerprint on the density field has multiple peaks and a much less trivial landscape, but would still represent a unique descriptor of the system properties. 

\section{ADDITIONAL FIGURES}

\begin{figure}[htb]
    \centering
    \includegraphics[width=1.0\textwidth]{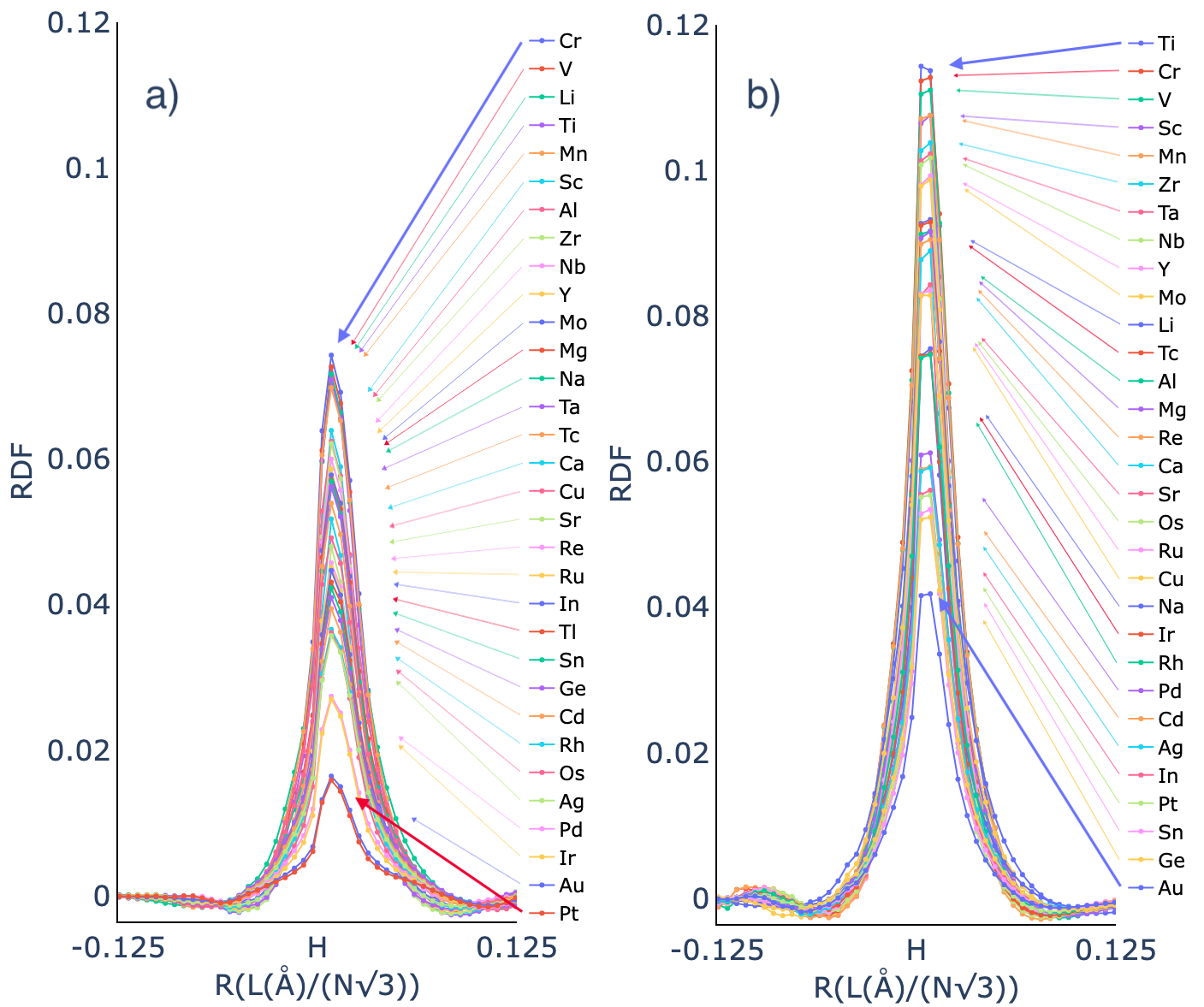}
    \caption{Density Radial Distribution Function profiles for the different metal species forming the host crystals, both for the OI case (a) and TI (b). The profiles are all centered around the relative hydrogen atom position in the grid box. Coloured arrows point towards the maximum of the corresponding RDF curve. The unique radius $R$ for the building of the RDFs comparison is here reported in units normalized by the number of grid points, $N=N_x=N_y=N_z=100$, and taking into account the three-dimensional nature of the distances. For completeness, some relevant quantities from the plots are reported in Tab.S.(III) and Tab.S.(IV) of SM.}
    \label{fig:CombinedRDFs}
\end{figure}

\begin{figure}[htb]
    \centering
    \includegraphics[width=1.0\textwidth]{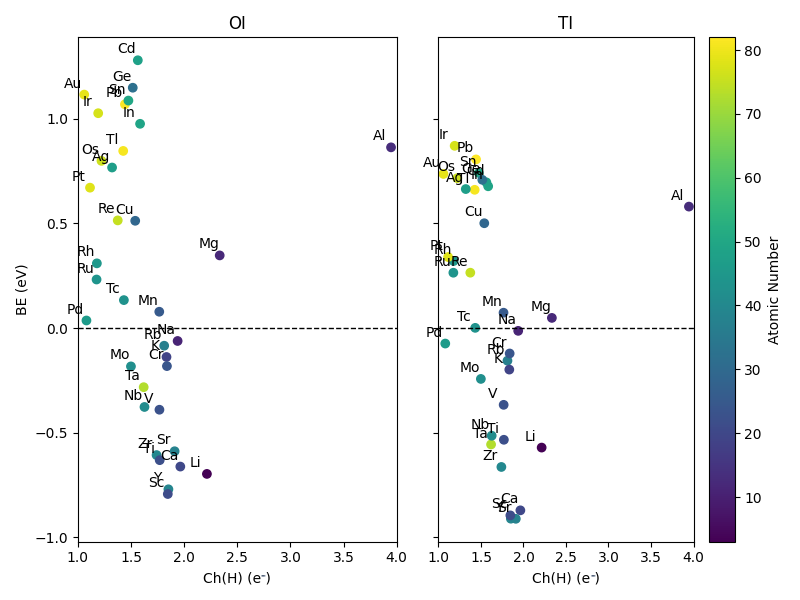}
    \caption{Visualization of the relation between the hydrogen binding energies and its Bader charge extensions in the different considered crystals.}
    \label{fig:BElinear}
\end{figure}

\section{TABLES}

\begin{table}[htb]
    \centering
\begin{tabular}{cccc} \toprule
      Crystal & $\text{E}_{\text{cut}}\text{(Ry)}$ &  $\text{k}_{\text{pts.}}$ &    $\text{degauss(Ry)}$ \\ \hline
  Thallium & 120 & 8 & 0.05\\
      Lead & 120 & 12  & 0.02 \\
       Tin & 100 & 16  &0.025 \\
  Rubidium & 80 & 10 & 0.025 \\
    Indium & 100 & 14 & 0.04\\
 Zirconium & 100 & 10  & 0.02 \\
   Yttrium & 80 & 10 & 0.02\\
  Tantalum & 120 & 10 & 0.04\\
 Strontium & 80 & 10 & 0.015\\
   Rhenium & 120 & 14 & 0.03\\
   Niobium & 120 & 8 & 0.05\\
    Sodium & 100 & 10  & 0.04 \\
 Potassium & 80 & 10 & 0.03 \\
   Cadmium & 80 & 16 & 0.03 \\
    Silver & 120 & 12 & 0.05 \\
   Rhodium & 120 & 14 & 0.03  \\
    Osmium & 120 & 16 & 0.03 \\
Molybdenum & 120 & 12 & 0.001 \\
  Scandium & 100 & 8  & 0.03 \\
 Magnesium & 100 & 10 & 0.04 \\
  Vanadium & 100 & 10 & 0.05 \\
  Titanium & 100 & 8  & 0.03  \\
Technetium & 80 & 10 & 0.03 \\
 Ruthenium & 100 & 14 & 0.03  \\
  Platinum & 80 & 14 & 0.03 \\
 Palladium & 80 & 12 & 0.03  \\
   Iridium & 80 & 16 & 0.03 \\
 Germanium & 80 & 16  & 0.03 \\
  Chromium & 100 & 8  & 0.005 \\
   Calcium & 80 & 10 & 0.01 \\
      Gold & 80 & 14 & 0.001\\
  Aluminum & 80  & 10  & 0.04 \\
 Manganese & 100 & 8  & 0.03 \\
    Copper & 80 & 12  & 0.05 \\
   Lithium & 80 & 10 & 0.03 \\
   \hline
\end{tabular}
    \caption{Table with the converged DFT parameters implemented in the simulations of the pure FCC metallic samples. The kinetic energy cutoff for the charge density and the potential is taken as  $\text{E}_{\text{cut}}^{\text{rho}}=6\cdot\text{E}_{\text{cut}}$, while the number of k-points in the table is intended to represent the selected 3-dimensional mesh $\text{k}_{\text{pts.}}\times\text{k}_{\text{pts.}}\times\text{k}_{\text{pts.}}$ in k-space. During the convergence tests, the common acceptance threshold in the variation of the total energy upon parameter change has been set to $10^{-5}$ Ry.}
    \label{tab:optim}
\end{table}

\begin{table}[]
    \centering
\begin{tabular}{cccccc} \toprule
      Crystal & Lattice Const. M (\AA) &  B(kbar) &  Poisson R. & Young M. & Shear M.\\ \hline
  Thallium & 5.002120 &  296.44047 &   0.40522 &    41.28377 &    14.68947 \\
      Lead & 5.069333 &  407.73666 &   0.78395 &  -233.36752 &   -65.40753 \\
       Tin & 4.872865 &  439.46883 &   0.40446 &   216.13070 &    76.94449 \\
  Rubidium & 6.896321 &   33.59768 &   0.34958 &    26.82293 &     9.93748 \\
    Indium & 4.824002 &  343.24574 &   0.43818 &    95.77966 &    33.29902 \\
 Zirconium & 4.543623 &  930.20851 &   0.34802 &   823.87718 &   305.58726 \\
   Yttrium & 5.045803 &  392.16302 &   0.25913 &   562.49297 &   223.36647 \\
  Tantalum & 4.240945 & 1740.99067 &   0.09708 &  2258.63998 &  1029.38703 \\
 Strontium & 5.934284 &  130.09873 &   0.27180 &   168.15819 &    66.11013 \\
   Rhenium & 3.936890 & 3501.73236 &   0.28170 &  4582.32578 &  1787.59034 \\
   Niobium & 4.242892 & 1620.01587 &   0.66029 & -1462.08151 &  -440.30960 \\
    Sodium & 5.210005 &   84.89276 &   0.32649 &    77.96226 &    29.38663 \\
 Potassium & 6.466959 &   41.39434 &   0.33615 &    35.33335 &    13.22206 \\
   Cadmium & 4.525227 &  436.67101 &   0.66442 &  -131.12809 &   -39.39151 \\
    Silver & 4.183577 &  913.21303 &   0.37763 &   660.02634 &   239.55201 \\
   Rhodium & 3.865649 & 2436.38085 &   0.25757 &  3541.13412 &  1407.92280 \\
    Osmium & 3.884664 & 3698.84839 &   0.23318 &  5906.41516 &  2394.79156 \\
Molybdenum & 4.033093 & 2304.21667 &   0.79723 & -3992.55717 & -1110.75225 \\
  Scandium & 4.596587 &  506.92217 &   0.29678 &   608.82277 &   234.74411 \\
 Magnesium & 4.493598 &  365.42375 &   0.26644 &   505.05471 &   199.40004 \\
  Vanadium & 3.832294 & 1757.16501 &   0.81350 &  -211.67221 &   -58.36018 \\
  Titanium & 4.116707 & 1097.47681 &   0.36308 &   857.83362 &   314.66815 \\
Technetium & 3.902914 & 2850.00899 &   0.30875 &  3263.65295 &  1246.86241 \\
 Ruthenium & 3.844112 & 2884.20345 &   0.25197 &  4280.15149 &  1709.37112 \\
  Platinum & 4.026684 & 2280.07686 &   0.37831 &  1662.58253 &   603.12330 \\
 Palladium & 3.988079 & 1627.68208 &   0.36285 &  1322.69354 &   485.26651 \\
   Iridium & 3.917226 & 3231.09305 &   0.23926 &  5053.83387 &  2039.05015 \\
 Germanium & 4.352847 &  526.74907 &   0.41439 &   217.70530 &    76.96105 \\
  Chromium & 3.640683 & 2416.34647 &   0.94407 & -6413.44601 & -1649.49145 \\
   Calcium & 5.446610 &  186.12162 &   0.29113 &   229.70885 &    88.95667 \\
      Gold & 4.207946 & 1246.19358 &   0.35609 &   181.33488 &    66.85944 \\
  Aluminum & 4.051156 &  753.26295 &  -2.42950 &  6628.79311 & -2318.56397 \\
 Manganese & 3.524944 & 2610.34410 &   0.24555 &  3979.29705 &  1597.40238 \\
    Copper & 3.644979 & 1713.15436 &   0.36665 &  1364.14460 &   499.08408 \\
   Lithium & 4.288954 &  148.32103 &   0.34126 &   128.74396 &    47.99367 \\
   \hline
\end{tabular}
    \caption{The structural information of the pure bulk metals. The Bulk Moduli (B), Poisson Ratios, Young Moduli and Shear Moduli have been computed via Thermo\_PW in Quantum Espresso.}
    \label{tab:stucttable_OI}
\end{table}



\begin{table}[]
    \centering
\begin{tabular}{cccccccccc}
\toprule
      Crystal & VALENCE & Ch(M) ex.($\text{e}^{-}$) & Ch(H) ex.($\text{e}^{-}$) & R(M)(\AA) & R(H)(\AA) & V(M)(\AA$^3$) & V(H)(\AA$^3$) & RDF peak width & RDF MAX\\
\hline
  Thallium &  14.0002 &        1.090700 &        0.426291 & 2.887141 & 1.810707 & 202.957874 &  52.687930 &    5.123690 & 0.043096 \\
      Lead &  14.9998 &        1.092029 &        0.442550 & 2.925526 & 1.809477 & 212.174845 &  50.747294 &    3.311369 & 0.000320 \\
       Tin &  14.9998 &        1.116130 &        0.475228 & 2.754140 & 1.790510 & 186.371487 &  47.363754 &    5.076671 & 0.042318 \\
  Rubidium &   9.9999 &        1.169677 &        0.811846 & 3.233507 & 2.579271 & 481.020725 & 172.325983 &    2.308308 & 0.000124 \\
    Indium &  14.0002 &        1.114555 &        0.584974 & 2.745684 & 1.772557 & 180.425142 &  59.032840 &    5.124412 & 0.044712 \\
 Zirconium &  13.0001 &        0.995388 &        0.740299 & 2.443752 & 1.788788 & 157.555295 &  43.461600 &    4.817026 & 0.062148 \\
   Yttrium &  11.9998 &        1.030338 &        0.851696 & 2.619778 & 1.986491 & 212.747840 &  65.069771 &    4.632992 & 0.058647 \\
  Tantalum &  28.0003 &        1.000756 &        0.619039 & 2.409388 & 1.558314 & 128.334835 &  30.065419 &    4.868236 & 0.056160 \\
 Strontium &  11.0000 &        1.113777 &        0.910118 & 2.928166 & 2.383004 & 333.896394 & 117.093181 &    4.451627 & 0.047993 \\
   Rhenium &  15.9999 &        1.065569 &        0.374854 & 2.299586 & 1.320261 & 100.843211 &  19.611020 &    4.959082 & 0.045747 \\
   Niobium &  14.0001 &        1.083274 &        0.626436 & 2.412199 & 1.559032 & 125.868122 &  31.868896 &    4.942415 & 0.060012 \\
    Sodium &   9.9999 &        1.198209 &        0.938265 & 2.251995 & 2.338292 & 200.104177 & 152.026014 &    4.928493 & 0.056993 \\
 Potassium &  10.0001 &        1.190132 &        0.833606 & 2.908772 & 2.579914 & 388.794096 & 168.589937 &    3.080109 & 0.000342 \\
   Cadmium &  13.0001 &        1.109309 &        0.563714 & 2.571604 & 1.710287 & 149.213605 &  54.083418 &    5.149464 & 0.039453 \\
    Silver &  12.0000 &        1.042793 &        0.321617 & 2.490956 & 1.427438 & 120.711034 &  33.179946 &    5.376127 & 0.035751 \\
   Rhodium &  18.0001 &        0.999350 &        0.178495 & 2.349289 & 1.255547 &  96.555521 &  18.211878 &    5.245036 & 0.036602 \\
    Osmium &  17.0001 &        1.014469 &        0.221748 & 2.379844 & 1.261711 &  97.883616 &  16.598398 &    5.056293 & 0.036162 \\
Molybdenum &  15.0001 &        1.127255 &        0.499084 & 2.293173 & 1.481930 & 106.406886 &  24.481625 &    5.053025 & 0.057802 \\
  Scandium &  11.9998 &        1.208514 &        0.845836 & 2.400477 & 1.809642 & 150.205482 &  57.297644 &    4.871119 & 0.063933 \\
 Magnesium &  11.0002 &        1.207869 &        1.334277 & 1.964853 & 2.122914 & 138.399775 & 101.149970 &    5.170782 & 0.057169 \\
  Vanadium &  14.0002 &        1.004913 &        0.767163 & 2.024360 & 1.536250 &  94.678433 &  32.810298 &    5.229013 & 0.072671 \\
  Titanium &  13.0002 &        1.031297 &        0.769750 & 2.131976 & 1.649858 & 115.736830 &  39.986057 &    5.090211 & 0.070894 \\
Technetium &  16.0001 &        1.038577 &        0.432690 & 2.285423 & 1.365815 &  98.687362 &  21.985729 &    5.014717 & 0.053937 \\
 Ruthenium &  17.0001 &        0.999019 &        0.175542 & 2.389375 & 1.210719 &  95.050740 &  16.939753 &    5.184028 & 0.045108 \\
  Platinum &  10.9999 &        0.985363 &        0.113874 & 2.467398 & 1.268213 & 109.542614 &  19.348826 &    4.794035 & 0.015905 \\
 Palladium &  10.9999 &        0.979355 &        0.080380 & 2.418034 & 1.256054 & 106.356794 &  19.482662 &    5.118990 & 0.027426 \\
   Iridium &  10.0000 &        1.014016 &        0.190940 & 2.406424 & 1.233746 & 100.230725 &  17.751418 &    4.917427 & 0.027035 \\
 Germanium &  15.0000 &        1.121465 &        0.516142 & 2.466538 & 1.645134 & 132.218509 &  44.834682 &    5.305638 & 0.040993 \\
  Chromium &  14.9997 &        1.108383 &        0.837709 & 1.999145 & 1.401443 &  78.341578 &  29.980409 &    5.330796 & 0.074272 \\
   Calcium &  11.0002 &        1.070980 &        0.963679 & 2.604790 & 2.287240 & 262.507311 & 109.599336 &    4.575987 & 0.051780 \\
      Gold &  12.0002 &        1.023319 &        0.059576 & 2.605179 & 1.325308 & 123.650005 &  22.275065 &    5.052181 & 0.016455 \\
  Aluminum &   3.9999 &        1.476572 &        2.946357 & 0.703848 & 2.479957 &  93.710980 & 119.706033 &    5.284540 & 0.062454 \\
 Manganese &  15.9997 &        1.099465 &        0.765646 & 1.957536 & 1.356901 &  71.415206 &  26.929691 &    5.461539 & 0.069813 \\
    Copper &  11.9998 &        1.061955 &        0.539289 & 2.005434 & 1.377594 &  79.417147 &  30.107248 &    5.477359 & 0.049192 \\
   Lithium &   3.9999 &        1.778538 &        1.213930 & 1.544476 & 2.161298 &  36.699317 & 133.519060 &    5.296808 & 0.071685 \\
\hline
\end{tabular}
    \caption{The density information coming from the Bader Charge and RDF Analysis on the OI-defected supercell samples.}
    \label{tab:Bader_OI}
\end{table}

\begin{table}[]
    \centering
\begin{tabular}{cccccccccc}
\toprule
      Crystal & VALENCE & Ch(M) ex.($\text{e}^{-}$) & Ch(H) ex.($\text{e}^{-}$) & R(M)(\AA) & R(H)(\AA) & V(M)(\AA$^3$) & V(H)(\AA$^3$) & RDF peak width & RDF MAX\\
\hline
  Thallium &  14.0002 &        1.093567 &        0.356148 & 2.545377 & 1.465705 & 203.888848 &  48.841286 &    1.151142 & 0.000797 \\
      Lead &  14.9998 &        1.140725 &        0.413141 & 2.584940 & 1.508106 & 210.276691 &  50.398776 &    1.110279 & 0.000782 \\
       Tin &  14.9998 &        1.146056 &        0.447967 & 2.450224 & 1.480478 & 185.886156 &  49.321816 &    4.709087 & 0.053449 \\
  Rubidium &   9.9999 &        1.348649 &        0.778730 & 2.809091 & 2.040784 & 419.691058 & 165.045977 &    1.382962 & 0.000286 \\
    Indium &  14.0002 &        1.136333 &        0.450868 & 2.389979 & 1.465630 & 180.368311 &  50.313394 &    4.759745 & 0.056079 \\
 Zirconium &  13.0001 &        1.159861 &        0.787938 & 2.190511 & 1.311776 & 152.153758 &  41.626603 &    4.696433 & 0.103792 \\
   Yttrium &  11.9998 &        1.180116 &        0.836165 & 2.271268 & 1.456762 & 205.789103 &  56.499096 &    4.305434 & 0.099258 \\
  Tantalum &  28.0003 &        1.056152 &        0.705935 & 2.123984 & 1.224397 & 127.370855 &  31.141105 &    4.842529 & 0.102275 \\
 Strontium &  11.0000 &        1.269420 &        0.869590 & 2.516584 & 1.699554 & 312.108926 &  95.041288 &    3.878054 & 0.084323 \\
   Rhenium &  15.9999 &        1.113671 &        0.495037 & 2.068914 & 1.077999 & 100.422245 &  21.926330 &    5.064890 & 0.090493 \\
   Niobium &  14.0001 &        1.088275 &        0.684128 & 2.120602 & 1.224953 & 126.613851 &  32.139568 &    4.896473 & 0.101722 \\
    Sodium &   9.9999 &        1.463217 &        0.893035 & 1.943787 & 1.850117 & 145.656724 & 148.928045 &    4.479028 & 0.075516 \\
 Potassium &  10.0001 &        1.422602 &        0.806801 & 2.539632 & 1.999094 & 311.017251 & 166.239056 &    1.172665 & 0.000790 \\
   Cadmium &  13.0001 &        1.102413 &        0.391133 & 2.251143 & 1.321713 & 150.053184 &  41.895034 &    5.018768 & 0.059318 \\
    Silver &  12.0000 &        1.060015 &        0.248673 & 2.149160 & 1.141106 & 120.659403 &  27.813010 &    5.193334 & 0.059168 \\
   Rhodium &  18.0001 &        1.023506 &        0.238498 & 2.019039 & 0.988487 &  96.107919 &  18.714500 &    5.284418 & 0.074714 \\
    Osmium &  17.0001 &        1.077913 &        0.389592 & 2.083358 & 0.993345 &  96.824240 &  19.239292 &    5.083399 & 0.083542 \\
Molybdenum &  15.0001 &        1.110887 &        0.565290 & 2.097791 & 1.100053 & 108.221570 &  25.653460 &    5.010381 & 0.098712 \\
  Scandium &  11.9998 &        1.062693 &        0.879424 & 2.007437 & 1.396542 & 160.374054 &  52.697116 &    4.722296 & 0.107582 \\
 Magnesium &  11.0002 &        1.314572 &        1.097658 & 1.699838 & 1.705034 & 131.015764 &  82.999180 &    4.963581 & 0.091603 \\
  Vanadium &  14.0002 &        1.013847 &        0.650536 & 1.869766 & 1.119319 &  94.868342 &  29.128173 &    5.442587 & 0.111030 \\
  Titanium &  13.0002 &        1.148948 &        0.800109 & 1.893797 & 1.212688 & 112.279150 &  38.441382 &    5.067758 & 0.114307 \\
Technetium &  16.0001 &        1.078196 &        0.423641 & 2.005951 & 1.064550 &  98.172688 &  22.278736 &    5.045718 & 0.092920 \\
 Ruthenium &  17.0001 &        1.061316 &        0.339811 & 2.047080 & 0.983592 &  93.881196 &  19.946138 &    5.250330 & 0.083500 \\
  Platinum &  10.9999 &        1.041885 &        0.192874 & 2.131596 & 0.942638 & 108.417780 &  19.844089 &    4.839737 & 0.055412 \\
 Palladium &  10.9999 &        1.042516 &        0.213853 & 1.998600 & 1.013624 & 104.660523 &  21.921053 &    5.062282 & 0.061210 \\
   Iridium &  10.0000 &        1.054041 &        0.374941 & 2.008881 & 0.995622 &  99.705215 &  20.216678 &    4.879931 & 0.074878 \\
 Germanium &  15.0000 &        1.120860 &        0.355051 & 2.196150 & 1.350224 & 132.143784 &  39.080898 &    5.234845 & 0.052370 \\
  Chromium &  14.9997 &        1.094787 &        0.603630 & 1.767076 & 1.071638 &  79.731303 &  24.965407 &    5.461745 & 0.112767 \\
   Calcium &  11.0002 &        1.342536 &        0.916390 & 2.204972 & 1.654794 & 230.761276 &  89.605791 &    4.166815 & 0.088972 \\
      Gold &  12.0002 &        1.025995 &        0.133964 & 2.197147 & 1.076686 & 124.290160 &  23.280690 &    5.042500 & 0.041849 \\
  Aluminum &   3.9999 &        1.392441 &        1.738636 & 0.655547 & 2.392302 &  96.574625 &  80.546765 &    5.118417 & 0.091678 \\
 Manganese &  15.9997 &        1.098617 &        0.496781 & 1.751445 & 1.037575 &  71.685497 &  21.893630 &    5.599272 & 0.107534 \\
    Copper &  11.9998 &        1.092595 &        0.325503 & 1.769494 & 1.000538 &  78.698192 &  23.330764 &    5.554652 & 0.082863 \\
   Lithium &   3.9999 &        1.738300 &        1.168284 & 1.288919 & 1.637778 &  41.523050 & 129.875995 &    5.351738 & 0.093254 \\
\hline
\end{tabular}
    \caption{The density information coming from the Bader Charge and RDF Analysis on the TI-defected supercell samples.}
    \label{tab:Bader_TI}
\end{table}

\begin{table}[]
    \centering
\begin{tabular}{cc|c} \toprule
      Mo&$2\times2
      \times2\,\,(\text{e/bohr}^3$)&$3\times3\times3\,\,(\text{e/bohr}^3)$\\ \hline
     Green isol. & 0.005 & 0.005\\
     Yellow isol. & 0.01168 & 0.01168 \\
     Max isol. & 0.0804678 & 0.07499216 \\
     Min isol. & -0.00838115& -0.00889763\\
   \hline
\end{tabular}
    \caption{Isoline values for the size effects test reported in Fig.S.(1).}
    \label{tab:isol_supercells}
\end{table}

\begin{table}[]
    \centering
\begin{tabular}{ccc} \toprule
             Features &       PC1 &       PC2 \\
VALENCE & -0.703 &0.096 \\
Ch(M) exchange &0.857	&-0.251\\
Ch(H) exchange &0.885	&-0.165\\
Radius of M charge &-0.506	&0.813\\
Radius of H charge&0.816	&0.362\\
Volume of M charge &0.261&0.904\\
Volume of H charge &0.922	&0.238\\
RDF peak width &-0.253	&-0.770\\
Max of RDF &0.243	&-0.440\\
   \hline
\end{tabular}
    \caption{Table of the computed loadings on PC1 and PC2 on the pure density dataset.}
    \label{tab:loads_densonly}
\end{table}